\documentclass[11pt,onecolumn,draftclsnofoot]{IEEEtran}
\usepackage{fancyhdr}
\usepackage{epsfig}
\usepackage{threeparttable}
\usepackage{epsf,epsfig}
\usepackage{amsthm}
\usepackage[cmex10]{amsmath}
\usepackage{amssymb, amsfonts}
\usepackage[noadjust]{cite}
\usepackage{dsfont}
\usepackage{subfigure}
\usepackage{color}
\usepackage{soul}
\usepackage{amssymb}

\newlength{\eqboxstorage}

\newtheorem{theorem}{Theorem}

\newtheorem{lemma}{Lemma}
\newtheorem{corollary}{Corollary}

\newtheorem{assumption}{Assumption}

\newtheorem{remark}{\bf Remark}

\def\qed{$\Box$}

\def\E{\mathsf{E}}

\def\phi{\varphi}

\def\SINR{\mathsf{SINR}}
\def\SNR{\mathsf{SNR}}
\def\SIR{\mathsf{SIR}}

\def\l{\left}
\def\r{\right}
\def\({\left(}
\def\){\right)}

\def\ba{{\mathbf{a}}}
\def\bb{{\mathbf{b}}}
\def\bc{{\mathbf{c}}}

\def\bee{{\mathbf{e}}}

\def\bs{{\mathbf{s}}}

\def\b0{{\mathbf{0}}}

\def\bQ{{\mathbf{Q}}}

\def\bV{{\mathbf{V}}}

\def\cJ{\mathcal{J}}

\newcommand{\nn}{\nonumber}

\DeclareMathAlphabet\mathbfcal{OMS}{cmsy}{b}{n}

\title{Communication Using a Large-Scale Array of Ubiquitous Antennas: A Geometry Approach}
\author{Kaibin Huang, Jiayi Chen and Vincent K. N. Lau    \thanks{\setlength{\baselineskip}{13pt} \noindent K. Huang is  with the Dept. of Electrical and Electronic Engineering at The  University of  Hong Kong, Hong Kong (Email: huangkb@eee.hku.hk). J. Chen is with the College of Information Engineering, Shenzhen University, China (Email:  i.c.jiayi@ieee.org).    V. K. N.  Lau  is with the Dept. of Electrical and Computer Engineering at  The Hong Kong University of Science and Technology, Hong Kong (Email: eeknlau@ece.ust.hk). Updated on \today.}}

\begin{document}

\maketitle

\begin{abstract} 
The recent trends of densification and centralized signal processing in radio access networks suggest that future networks may comprise ubiquitous antennas coordinated   to form a network-wide  gigantic array, referred to as the \emph{ubiquitous array} (UA).  In this  paper, the  UA communication techniques are designed and analyzed based on a geometric model. Specifically,  the UA is modeled as a continuous circular/spherical array enclosing target users  and free-space propagation is assumed. First, consider the estimation of multiuser  UA channels induced by user locations. Given single pilot symbols, a novel channel estimation scheme is proposed that decomposes training signals into Fourier/Laplace  series and thereby  translates multiuser channel estimation into peak detection of a derive function of location. The process is shown  to suppress noise.  Moreover, it is proved that estimation error due to interference diminishes with the increasing minimum user-separation distance  following the \emph{power law}, where the exponent is $1/3$ and $1$ for the circular and spherical UA, respectively. If orthogonal pilot sequences are used, channel estimation is found to be perfect.  Next, consider channel-conjugate data transmission that maximizes received signal power.  The power of interference between two users is shown to decay with the increasing user-separation distance \emph{sub-linearly} and \emph{super-linearly} for the circular and spherical UA, respectively. Furthermore, a novel  multiuser precoding design is proposed by exciting  different  phase modes of the UA and controlling the mode weight factors to null interference. The number of available degrees of freedom for interference nulling using the UA is proved to be proportional to the minimum user-separation distance. 

\end{abstract}

\section{Introduction}
The explosive growth of mobile traffic is driving the rapid network densification  to provide high-speed wireless access to coverage regions. To rein in the escalating network cost and interference, wireless access networks are evolving towards having an architecture with centralized signal processing and minimum on-site hardware comprising merely antennas and RF units, called either cloud radio access networks or base-station  virtualization \cite{C-RAN}. The network evolution as well as the advancements of other  technologies such as small cells \cite{ChanAndrews:FemtocellSurvey:2008} and massive MIMO \cite{RusekLarssonMarz:ScaleUpMIMO:2012} will lead to future networks where antennas are ubiquitous and have network-wide coordination to form a gigantic array, which is called the \emph{ubiquitous array} (UA) and forms the theme for the paper. 

\subsection{Prior Work and New Challenges}

The UA system is equivalent to a \emph{distributed antennas system} (DAS) with dense antennas and without cells. DASs  refers to  cellular systems where in each cell, antennas  are distributed over the cell region and connected to a central processing unit \cite{Heath:DistributedAntennas:2013}. This technology was initially developed to reduce transmission power and improve network coverage by either simulcast over all distributed antennas or serve each user  using the nearest antennas \cite{Choi:DownlinkDASMultiCell:2007, Zhang:DASwithRandomAntennas:2008}. Recent research on DASs focuses on increasing the sum rate based on multiuser  MIMO transmission  using the distributed antennas as a virtual array, addressing issues such as inter-cell interference distribution \cite{Heath:DistributedAntennasSysOutOfCellInterference:2011}, resource allocation \cite{Zhu:RadioResourceAllocationDAS:2013}, capacity analysis \cite{KKWong:LargeScaleMACDistributedAnt:2013, McKay:SumRateAnalysisZFRXDistributedMIMO:2013, Lin:CompareCoLocatDistributedAntennas:2014} and multi-cell coordination \cite{Feng:VirtualMIMODistributedAntCoordTX:2013}.  Each MIMO channel in such systems has coefficients corresponding to heterogeneous path loss depending on antenna locations and cannot be modeled as i.i.d. random variable as for the case of co-located antennas (see e.g., \cite{Tel:CapaMultGausChan:99}). This complicates the distributions of signals and interference  \cite{Heath:DistributedAntennasSysOutOfCellInterference:2011} and provides extra degrees of freedom, namely antenna locations, for sum-rate optimization \cite{Wang:AntennaLocDAS:2009, ParkLee:AntennaPlacementDistributedAntenna:2012}.  Essentially, this work is an attempt to quantify the advantages of extremely dense distributed antennas for channel estimation and data transmission.

Given the proximity between the UA and users, the UA channel is  typically  over free space or at most contains sparse scatterers. Combining free space channels and ubiquitous antennas shifts the paradigm of MIMO communications in several aspects.   First, a new approach is needed for analyzing the capacity of the UA channel. Rich scattering is commonly assumed in a conventional MIMO channel, allowing the channel to be modeled as a random matrix and its capacity analyzed using probability theory and linear algebra (see e.g., \cite{Tel:CapaMultGausChan:99}). This approach, however, is unsuitable for the UA channel  with free space propagation since the channel capacity depends on the array geometry and the user locations. Thus, analyzing the capacity of the UA channel should rely on an approach merging geometry, electromagnetic wave theory and information theory in the same vein as \cite{PoonTse:DoFMultiAntennaChannels:2005, Gruber:EMInfoTheoryWireless:2008}. Next, despite the massive number of elements in the UA, the UA channel over free space   is determined only by a few parameters  such as the user locations and this fact can be exploited to dramatically reduce the complexity of channel estimation. In contrast, given rich scattering, deploying more antennas leads to continuous growth of the number of degrees of freedom (DoF) in the channel, which makes channel estimation a key challenge in designing massive MIMO systems \cite{Marzetta:CellularUnlimitedBSAntennas:2010}. Last, the classic technique  for multiuser beamforming for free space channels computes nulls by solving a linear system where the number of variables is equal to that of transmit antennas \cite{VanVeen:Beamforming:1988}, and thus is inefficient for the UA system with a massive number of transmit antennas. This calls for the design of efficient transmission algorithms for the UA systems.

\subsection{Contributions and Organization}

The paper represents the  first attempt on designing  the UA communication systems and focuses on the signal-processing aspect, namely channel estimation and data transmission. For tractability, the work considers a particular coverage region represented by  a simple geometric model which comprises a continuous circular  UA communicating with $U$ single-antenna users at  fixed locations near the UA center. In the model, propagation is constrained to be within the horizontal plane. The  results are subsequently  extended to the case  with propagation in  the three-dimensional space and a  continuous spherical UA. Channels are assumed to be free space, narrow band, and reciprocal. The elements  of the UA and user antennas   are all assumed to be  omnidirectional. The layout of the above model is similar to that of some existing ones for single-cell DASs (see e.g.,  \cite{Yang:MassiveMIMOCircularlyDistributedAntennas}) that, however, assumes discrete  antennas and rich scattering. The continuity of the UA, modeling dense antennas, is a typical technique for avoiding consideration of antenna placement (see e.g., \cite{PoonTse:DoFMultiAntennaChannels:2005}). More important, it is  instrumental for the new findings and the algorithmic designs  as summarized in the sequel.

First, consider estimation of multiuser UA channels using only single pilot symbols. The channels  are determined by user locations and thus called \emph{location induced channels} (LI-channels). The channel responses are non-linear functions of the  locations. This makes  the conventional linear (mimimm-mean-square-error or zero-forcing)  estimation  unsuitable and the optimal maximum-a-posteriori estimation intractable since it requires  solving a set of non-linear equations \cite{VanTrees:DetectionEstimation:2004}.  To address this issue, a novel low-complexity channel-estimation technique  is proposed based on decomposing the receive multiuser circular training signal into a Fourier series for the circular UA or spherical harmonics for the spherical UA. This leads to a derived  function of location, called the \emph{user-location profile}. The  proposed estimation method  is to detect the locations of the peaks of the profile that yield  estimated user locations. The estimation procedure is shown to suppress noise by averaging and incur estimation errors only due to interference between multi-channel etimation. The error is shown to decay with the minimum user-separation distance following the power law with the exponent $1/3$ and $1$ for the circular and spherical UA, respectively.  Therefore, without orthogonal pilot sequences, multiuser channel estimation in the UA system is enabled by sufficiently large user-separation  distances  instead of differentiation  in multiuser angles of arrival as in the conventional MIMO systems (see e.g.,  \cite{YinGesbert:CoordinatedApproachLargeScaleMIMO:2013}). In addition, applying the method to the scenario where users deploy orthogonal pilot sequences leads to almost-perfect channel estimation.

Next, consider channel conjugate data transmission using the UA. The signal-to-interference-and-noise ratios (SINRs) are derived in closed-form. In particular, the power of interference between any two users is shown be proportional to their  separation distance (in wavelength) raised to the power of $2/3$ and $2$, corresponding to the circular and spherical UA, respectively.   Therefore, even given single-user transmission, interference can be suppressed by increasing users' separation distances. Moreover, the path lose is shown to be inversely proportional to the propagation distance or \emph{fixed} regardless of the distance for the circular and spherical UAs, respectively. In contrast,  the loss  for a conventional array with collocated elements is inversely proportional to the \emph{squared} distance. 

Last,  a novel low-complexity  precoding design  is proposed for multiuser transmission using the UA. Specifically, the precoders are designed in the form of Fourier series for the circular UA and spherical harmonics for the spherical UA. Their coefficients are controlled as derived  to excite different \emph{phase modes} of the circular array so as to null multiuser interference. For this sub-optimal design, the number of DoF available for interference nulling is shown to be proportional to the minimum user separation distance  or its square for the circular and spherical UAs, respectively. 

The reminder of the paper is organized as follows. The UA system model is described  in Section~\ref{Section:System}.  Algorithms for channel estimation and data transmission for the circular-UA system are presented in Section~\ref{Section:ChanEst} and \ref{Section:DataTx}, respectively. The results are extended to the spherical-UA system  in Section~\ref{Section:Spherical:UA}.  Simulation  results are provided in Section~\ref{Section:Sim} follows by concluding remarks in Section~\ref{Section:Conclusion}. In Appendix~\ref{App:Bessel}, Bessel functions and spherical harmonics are defined and their properties  discussed. Last, Appendix~\ref{App:Proofs} contains the proofs for lemmas.

\section{System Model}\label{Section:System}

\begin{figure}[t]
\centering
\subfigure[Circular-UA System ]{\includegraphics[width=7.5cm]{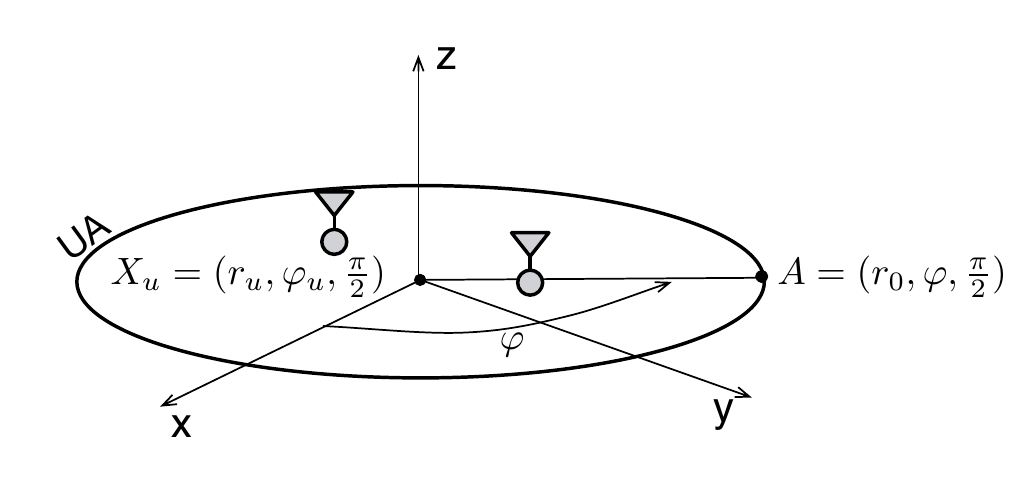}}
\subfigure[Spherical-UA System]{\includegraphics[width=8.5cm]{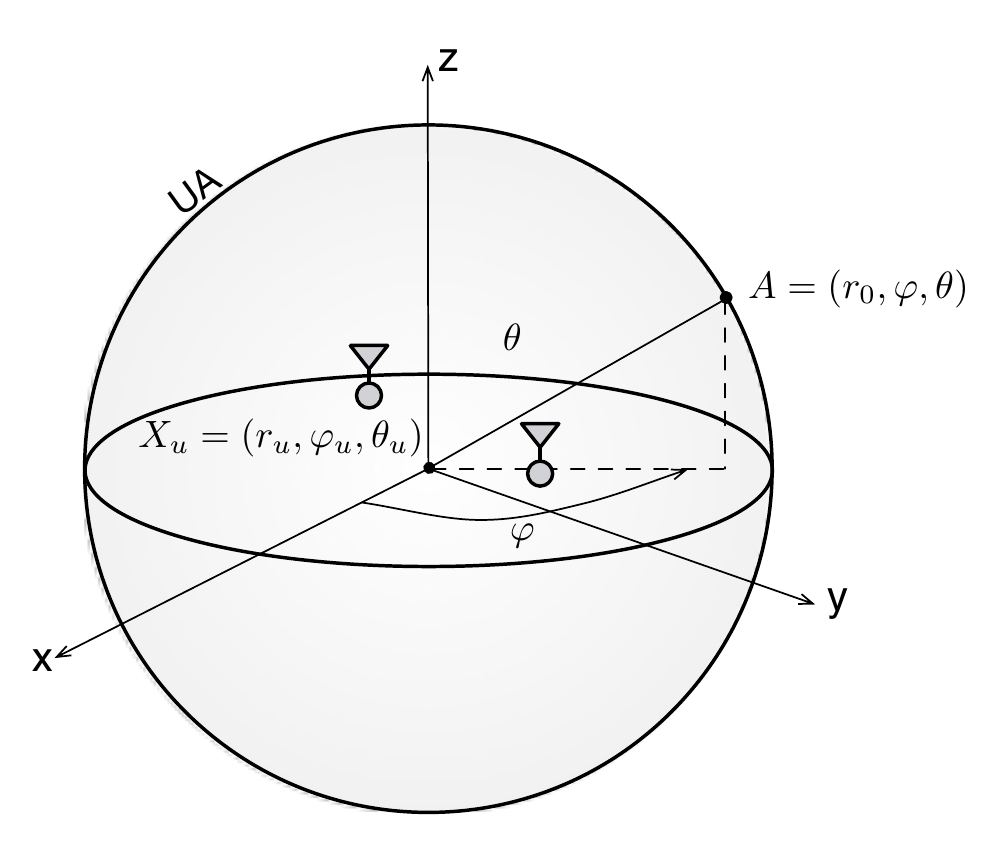}}
\caption{A geometric model of a communication system using a continuous circular/spherical UA.   }
\label{Fig:UASystem}
\end{figure}

As illustrated in Fig.~\ref{Fig:UASystem}, the UA is modeled as either a \emph{circular array} or a \emph{spherical array}, denoted as $\mathds{O}$ and $\mathds{O}^2$, respectively, having  the same radius denoted as $r_0$. The dense UA is assumed to be \emph{continuous} for tractable analysis but this assumption is relaxed in simulation.  The  communication system comprises  the UA centered at the origin   and a set of  $U$ single-antenna users   enclosed by the UA, represented by their fixed locations  $X_1, X_2, \cdots, X_U$ in the horizontal plane.  A  user,   $X_u$,  and a particular  element of the UA,  $A$,  are represented by their spherical coordinates $(r_u, \phi_u, \tfrac{\pi}{2})$ and $(r_0, \phi, \theta)$, respectively, where $(\phi_u, \phi)$ are azimuth angles and $(\tfrac{\pi}{2}, \theta)$ are polar angles with $\theta = \frac{\pi}{2}$ for the case of circular UA.  In addition, there are no scatterers. 

\begin{assumption}\label{AS:CenterUsers}\emph{Users are  located near the center of  the UA such that $r_u/r_0 \ll 1$ for $u = 1, \cdots, U$. }
\end{assumption}

\noindent The assumption allows tractable analysis as it  simplifies  the expression for the propagation distances. Specifically, given   a user  $X_u$ and a UA antenna  $A$, the separation distance and angle are denoted as $|X_u - A|$ and $\psi_u(A)$, respectively, with 
\begin{align}
|X_u - A| &= \sqrt{r_0^2 + r_u^2 - 2 r_0 r_u\cos\psi_u(A)}\nn\\
&=  r_0 - r_u\cos\psi_u(A) + o \label{Eq:Dist:Approx}
\end{align}
where $o$ represents  $O(\max_u r_u/r_0)$.  Note  that  the  above model can be extended to  include a set of scatterers at given locations, which reflect communication signals between the UA and users in  the absence of lines of sight. Remarks on the extensions of results to scattering channels are provided  in the sequel.

The wave transmitted by an  antenna  is  assumed to propagate as a plane wave in the three-dimensional free space. All antennas are assumed to be omni-directional.  Let $h_u(A)$ represents the response  of the channel between $X_u$ and $A$. As a result,   
\begin{equation}
 h_{u}(A) = \frac{1}{\sqrt{4\pi} |X_u - A|}e^{-j\frac{2\pi}{\lambda}|X_u - A|},\qquad A \in \mathds{O} \ \text{or} \ \mathds{O}^2\nn
\end{equation}
where $\lambda$ denotes the carrier wavelength. Based on   \eqref{Eq:Dist:Approx},  
\begin{equation}
 h_{u}(A)  =\frac{1}{\sqrt{4\pi} r_0} e^{-j\frac{2\pi }{\lambda} r_0 + j\frac{2\pi}{\lambda}r_u\psi_u(A)} + \frac{o}{r_0}, \qquad A \in \mathds{O} \ \text{or} \ \mathds{O}^2.   \label{Eq:Ch:Gain}
\end{equation}

Channel estimation at the UA is assisted by pilot signals transmitted by users. Time is divided into  slots with unit symbol  durations.  Since the effective aperture for  a omni-directional   antenna is $\lambda^2/4\pi$ \cite{Friis:TransmFormula:1946}, the total pilot signal received at antenna $A$ in an arbitrary slot, denoted as $q(A)$, is given as 
\begin{equation}\label{Eq:Sig:UL}
q(A)  =\frac{\lambda}{\sqrt{4\pi}}\sum_{u=1}^U  h_{u}(A)  s_{u}+ z(A), \qquad A \in \mathds{O}\ \text{or} \  \mathds{O}^2
\end{equation}
where $s_{u}$ is a pilot symbol transmitted by user $u$ and the noise  $z(A)$ is a spatial  sample of the additive white Gaussian noise  $\mathcal{CN}(0, \sigma^2)$ process at location $A$. The noise processes   are assumed to be \emph{spatially white} as follows. 
\begin{assumption}\label{AS:White}\emph{The noise processes  $z(X)$ and $z(Y)$ at two locations $X$ and $Y$ are independent if $X \neq Y$.}
\end{assumption}
\noindent Substituting \eqref{Eq:Ch:Gain} into \eqref{Eq:Sig:UL} gives 
\begin{equation}\label{Eq:Training:Sig}
  q(A) =\frac{\lambda e^{ -j\frac{2\pi}{\lambda}r_0}}{4\pi r_0}\sum_{u=1}^U e^{ j\frac{2\pi}{\lambda}r_u\psi_u(A)} s_{u}+ z(A) +\frac{o}{r_0}, \qquad A \in \mathds{O} \ \text{or} \  \mathds{O}^2.    
\end{equation}

Next, consider downlink data transmission. The data symbol  intended for user $u$ is  denoted as $x_u$ and assumed to be distributed as a $\mathcal{CN}(0, 1)$ random variable. The symbol is  precoded by a continuous precoder represented by  $f_u: \mathds{O} \ \text{or} \  \mathds{O}^2 \rightarrow \mathds{C}$. Let $P_{\textrm{t}}$ denote the transmission power per user.  Then the incident field at  location  $u$ in an arbitrary slot, denoted as $g(X_u)$,  can be written as 
\begin{equation}
g(X_u) =\sqrt{\frac{P_t}{\partial\mathcal{A}}} \int_{\mathcal{A}}h_u(A) \sum_{k=1}^U f_k(A) \ dA \ x_{k}, \qquad \mathcal{A} = \mathds{O} \ \text{or}\  \mathds{O}^2 \label{Eq:Sig:DL}  
\end{equation}
with $h_u(A)$ being  the channel response  in \eqref{Eq:Ch:Gain} and $\partial\mathcal{A}$ represents the circumference  of a circular UA ($\mathcal{A} = \mathds{O}$) or the surface area of a spherical UA ($\mathcal{A} = \mathds{O}^2$).   It follows that the corresponding received signal  is 
\begin{equation}
y_u = \frac{\lambda }{\sqrt{4\pi}}g(X_u) + z_u, \qquad u = 1, 2, \cdots, U
\end{equation}
where  $\{z_{u}\}$ are i.i.d. $\mathcal{CN}(0, \sigma^2)$ random variables representing channel noise.

\section{Communication Using the Circular UA: Channel Estimation}\label{Section:ChanEst}
Estimation of the LI-channels is to infer the user locations from the training signal, namely 
\begin{equation}
\{q(A)\mid A \in \mathds{O} \}\Longrightarrow \{X_u\}\nn
\end{equation}
with $\{q(A)\}$ given in \eqref{Eq:Training:Sig}. As mentioned, the linear or MAP estimation techniques are intractable  since $q(A)$ is a nonlinear function of $\{X_u\}$. 
To address this issue, a simple estimation scheme is  proposed in the following sub-sections, which reduces channel estimation to the detection of the peaks of a given function. 

\subsection{LI-Channel Estimation with  Single Pilot Symbols}\label{subsec:chn_estm_single_pilot}
Consider  the scenario where users  simultaneously  transmit \emph{single} pilot symbols $\{s_u\}$ to facilitate channel estimation at the UA.  Without loss of generality, assume that the pilot symbols are all ones: $s_u = 1\ \forall \ u$.    Let  the training signal $q(A)$ in \eqref{Eq:Sig:UL} be re-denoted as $q(\phi)$ since $A = (r_0, \phi, \tfrac{\pi}{2})$.

The  LI-channel estimation scheme is designed as follows.  First, the proposed   scheme  is based on decomposing  the received training signal $\{q(\phi)\}$  into  a  Fourier series: 
\begin{equation}
q(\phi) = \sum_{k = -\infty}^\infty Q_k e^{-j k\phi}, \qquad \phi \in [0, 2\pi]
\end{equation}
where the Fourier coefficients $\{Q_k\}$ are defined as 
\begin{equation} \label{Eq:Fourier}
Q_k  = \frac{1}{2\pi} \int_0^{2\pi} q(\phi)\ d\phi. 
\end{equation}
The Fourier coefficients contain all information on  the signal and thus can replace it in   channel estimation.  To facilitate the algorithmic design, the structure of the   coefficients is characterized as follows.  To this end, a few notations are introduced.  Let $\bQ$ represent the infinite sequence: $\cdots, Q_{-1}, Q_0, Q_1, \cdots$. The product between two sequences, $\bV_1$ and $\bV_2$,  is denoted and defined as 
$\bV_1\circ\bV_2 = \sum_{n=-\infty}^{\infty}[\bV_1]^*_n [\bV_2]_n$ where $[\cdot]_n$ yields the element with index $n$. Moreover, $J_n: \mathds{R}\rightarrow\mathds{R}$ represents Bessel function with an integer index $n$ as defined and discussed in Appendix~\ref{App:Bessel}.

\begin{lemma}[Training signal decomposition]\label{Lem:Fourier}The sequence $\bQ$ can be decomposed as 
\begin{equation}
\bQ =  \frac{\lambda}{4\pi r_0}e^{-j\tfrac{2\pi}{\lambda}r_0} \sum_{u=1}^U \bV(X_u) +  \frac{o}{r_0}, \qquad \text{a.s.}
\end{equation}
where $\bV(X_u) \!=\! \cdots V_{-1}(X_u), V_0(X_u), V_1(X_u) \cdots$ with the function  $V_n(Y)$ defined for a given location $Y= (r_Y, \phi_Y, \tfrac{\pi}{2})$ as 
\begin{equation}
V_n(Y) = j^n  e^{jn\varphi_Y} J_n\left(\tfrac{2\pi }{\lambda}r_Y\right).
\end{equation}
\end{lemma}
The proof is provided in  Appendix~\ref{Lem:Fourier:Proof}.  
\begin{remark}\emph{The Fourier coefficients $\{Q_k\}$ of the received training  signal are \emph{noiseless}. The noise suppression is the combined result of the noise spatial whiteness in Assumption~\ref{AS:White} and the integral operation  in \eqref{Eq:Fourier}.} 
\end{remark}

Next, based on the decomposition of $\bQ$ in Lemma~\ref{Lem:Fourier}, the key component of the proposed scheme  for LI-channel estimation is  a   function $\Phi: \mathds{R}^2 \rightarrow \mathds{R}_+$ defined as 
\begin{equation}\label{Eq:Loc:Profile}
\Phi(Y) = \frac{4\pi r_0 }{\lambda}|\bV(Y)\circ\bQ|
\end{equation}
and called the \emph{channel observation profile }. To derive a closed-form expression for $\Phi(Y)$, a useful property for  $\bV$ directly follows from the Addition Theorem in Property (B2) of  Bessel functions in Appendix~\ref{App:Bessel} as shown below. 
\begin{lemma}\label{Lem:Proj}Given two locations $X, Y\in \mathds{R}^2$, the product between the sequences $\bV(X)$ and $\bV(Y)$ satisfies 
\begin{equation}
|\bV(X)\circ\bV(Y)| = J_0\left(\tfrac{2\pi }{\lambda}|X- Y|\right). \nn
\end{equation}
\end{lemma}
Combining Lemmas~\ref{Lem:Fourier} and \ref{Lem:Proj} and the definition of $\Phi(Y)$ in \eqref{Eq:Loc:Profile} gives the following theorem. 

\begin{theorem}[Channel observation] \label{Theo:Profile}The channel observation profile  corresponding to the circular UA is noiseless and given as 
\begin{equation} \label{Eq:Loc:Profile:a}
  \Phi(Y) = \left |\sum\nolimits_{u=1}^U J_0\left(\tfrac{2\pi }{\lambda}|X_u - Y|\right)\right|  + o, \quad \text{a.s.}  
\end{equation}
\end{theorem}

\begin{figure}[t]
\begin{center}
\includegraphics[width=9cm]{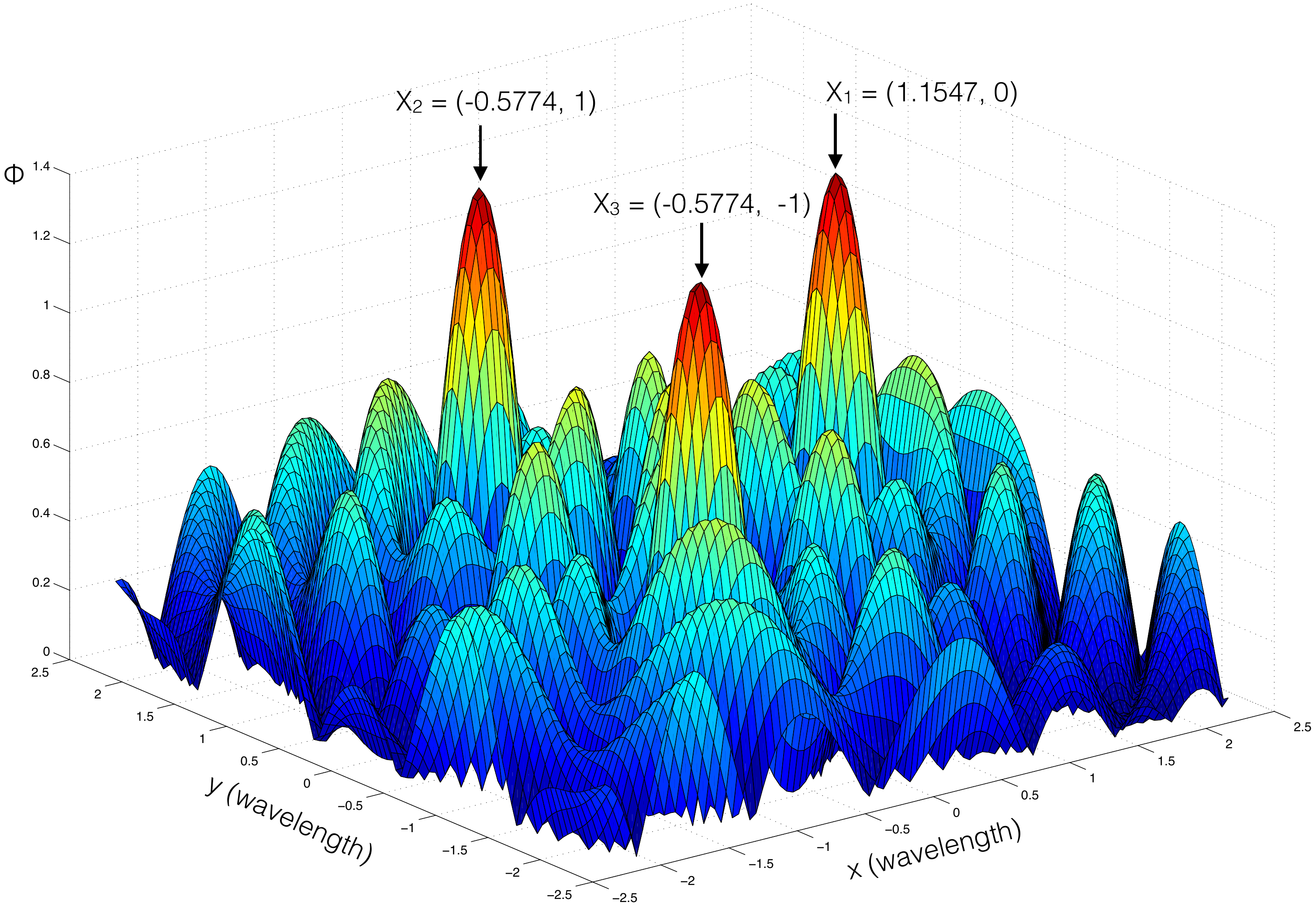}\vspace{10pt}
\caption{ An example of the channel observation profile  corresponding to three users with  equal separation distances of $2$   wavelengths where the peaks are identified by their Cartesian coordinates and the ripples arise from the tails of superimposed Bessel functions (see Theorem~\ref{Theo:Profile}). }
\label{Fig:Profile}
\end{center}
\end{figure}

An example of $\Phi(Y)$ is illustrated in Fig.~\ref{Fig:Profile}. 
 
 {\bf LI-channel estimation scheme:} One can observe from Theorem~\ref{Theo:Profile} that the $U$ Bessel functions in $\Phi(Y)$ have their peaks at corresponding user locations since $J_0(x)$ is maximized at $x = 0$. Motivated by this fact,  the proposed scheme for estimating the user  locations is to detect the peaks in the channel observation profile  $\Phi(Y)$.  Since the profile is noiseless, the only source for estimation errors is the coupling (interference) between the Bessel functions in $\Phi(Y)$ due to finite separation distances between user locations.

\begin{remark}[Channel estimation error] \label{Re:EstErr} \emph{Channel estimation using the proposed scheme  is close to perfect for a single-user system since the corresponding channel observation profile   $\Phi(Y) \approx  J_0\left(\tfrac{2\pi }{\lambda}|X_1 - Y|\right) $ that  is maximized at $Y = X_1$ with $\Phi(X_1) \approx  1$. Next, consider the estimation of multiuser LI-channels. The accuracy for estimating the channel  corresponding to user $X_u$ can be evaluated by the difference between $\Phi(X_u)$ and the value of  $1$ for its single-user counterpart. This  measures the interference due to the presence of multiuser channels and can  be  obtained from Theorem~\ref{Theo:Profile} as follows: 
\begin{equation}\nn
  |\Phi(X_u) - 1| \leq  \left |\sum\nolimits_{k\neq u } J_0\left(\tfrac{2\pi }{\lambda}|X_k - X_u|\right)\right|  + o, \quad \text{a.s.}  
\end{equation}
 Based on Property  (B4) of Bessel functions in Appendix~\ref{App:Bessel},  
\begin{align}
  |\Phi(X_u) - 1| &\leq  \frac{1}{\nu} \sum_{k \neq u} \l(\tfrac{2\pi}{\lambda} |X_k - X_u|\r)^{-\frac{1}{3}} + o, \quad \text{a.s.}  \nn\\
  &  \leq \frac{U - 1}{\nu}  \l(\tfrac{2\pi}{\lambda} \min_{u \neq k} |X_k - X_u|\r)^{-\frac{1}{3}} + o, \quad \text{a.s.} \label{Eq:ErrBnd}
\end{align}
The result shows that the interference magnitude  diminishes with the  increasing minimum user-separation distance approximately following a sub-linear function. Given a pair of  users, setting the upper bound in \eqref{Eq:ErrBnd} to a small value e.g., $0.1$, a rule of thumb for the user-separation distance sufficiently large for accurate multiuser channel estimation can be computed as $77\lambda$, which is  $23$ m for  a carrier  of $1$ Gz and  $2.3$ m for $10$ Gz. 
}
\end{remark}

\begin{remark}[Effect of wavelength on channel estimation] \emph{One can infer  from \eqref{Eq:ErrBnd} that with user locations fixed, the accuracy of channel estimation can be improved by increasing the carrier  wavelength  $\lambda$. However, this leads to a denser UA in practice since its elements are required to be separated by no more than $\lambda/2$. 
} 
\end{remark}

\begin{remark}[Scattering Channels]\emph{Assuming the reflection at scatterers are isotropic, the channels between the UA and users are determined not only  by the scatterers'  locations but also by the channel coefficients that combine  the gains of the channels between scatterers and users and reflection attenuation at scatterers.  Estimation of the channels  can follow a two-phase procedure. First, the scatter locations can be estimated following the same scheme as discussed earlier for estimating user locations.   Next, resolving the scatter locations allows the estimation of the training signals reflected  by individual  scatterers. However, further estimating the individual channel coefficient between each pair of scatterer and user requires the use of pilot sequences and exploiting  their orthogonality. In other words, unlike that of free-space channels, channel estimation using single pilot symbols is infeasible for scattering channels.}
\end{remark}

\subsection{LI-Channel Estimation with  Pilot Sequences}
In this section, the results in the preceding section for single pilot  symbols are extended to the case with pilot  sequences. Let the pilot sequence for the $u$-th user be denoted as  $\bs_{u} = [s_{u}(1),\cdots, s_{u}(L)]^T$ with length $L$. In the channel training phase, the UA receives sequentially $L$ infinite sequences     over $L$ symbol durations,  denotes as $\bQ(1), \bQ(2), \cdots, \bQ(L)$,  where  $\bQ(\ell)$ is modified from its single-symbol counterpart  in Lemma~\ref{Lem:Fourier} as 
\begin{equation}
\bQ(\ell) = \frac{\lambda e^{-j\tfrac{2\pi}{\lambda}r_0}}{4\pi r_0} \sum_{u=1}^U s_u(\ell)\bV(X_u) +  \frac{o}{r_0}, \qquad \text{a.s.}\label{Eq:Q:ell}
\end{equation}
To estimate the user location $X_u$, $\{\bQ(\ell)\}$ are coherently  combined using  $\bs_u$ as $\sum_{\ell=1}^L s^*_{u}(\ell)\bQ(\ell)$. The result,  denoted as $\bQ_u$,  follows from \eqref{Eq:Q:ell} as 
\begin{equation}
\bQ_u =  \frac{\lambda e^{-j\tfrac{2\pi}{\lambda}r_0}}{4\pi r_0} \sum_{k=1}^U \bs^\dagger_u 
\bs_k\bV(X_k) +  \frac{o}{r_0}, \qquad \text{a.s.}
\end{equation}
A corresponding channel observation profile  $\Phi_u(Y)$ can be defined similarly as in \eqref{Eq:Loc:Profile}:  
\begin{equation}\label{Eq:Loc:Profile:Seq}
\Phi_u(Y) = \frac{4\pi r_0}{\lambda}|\bV(Y)\circ\bQ_u|. 
\end{equation}
The profile $\Phi_u(Y)$ is  decomposed into 
the desired and interference terms as follows: 
\begin{align}
\Phi_u(Y) \!=\! \Big |  J_0\left(\tfrac{2\pi}{\lambda}|X_{u}\!-\!Y|\r) \!+\! &\sum_{k\neq u} \bs^\dag_{u}\bs_{k} J_0\left(\tfrac{2\pi}{\lambda}|X_k\!-\! Y|\r)\Big |^2 \!+\!o.  \nn
\end{align}
Then the deviation of $\Phi_u(Y)$ from its single-user counterpart  can been  bounded as: 
\begin{equation}
\l|\Phi_u(Y)-J_0^2\left(\tfrac{2\pi }{\lambda}|X_{u}-Y|\right)\r| \leqslant \sum\nolimits_{k\neq u} |\bs^\dag_{u}\bs_{k}| J_0\left(\tfrac{2\pi}{\lambda}|X_k -Y|\right)+o, \qquad \forall \ u. \label{Eq:Loc:Profile:Seq:UB}
\end{equation}
This leads to the following main result of this section. 
\begin{theorem}[Effect of pilot sequences] \label{Theo:Profile:Seq}
Given $L \geq U$ and orthogonal pilot sequences, the LI-channel estimation using the circular  UA is almost perfect since the channel observation profile  is approximately equal to the single-user counterpart:
\begin{align}
\Big |\Phi_u(Y) - & J_0\left(\tfrac{2\pi}{\lambda}|X_{u} - Y|\right) \Big |\leqslant  o, \qquad u =1, 2, \cdots, U. \nn
\end{align}
\end{theorem}

%
Comparing Theorems~\ref{Theo:Profile} and \ref{Theo:Profile:Seq}, the advantage   of pilot sequences over single pilot symbols lies in their capability of  decoupling the estimation of multiuser LI-channels, thereby providing close-to-perfect channel estimation. 

\subsection{Effect of Finite Elements in the UA}

Consider a discrete circular UA with $N$ antennas uniformly placed on the circle centered at the origin and with the radius $r_0$. The discrete UA can be interpreted as a quantized version of the continuous UA with the quantization error bounded by $\pi/N$. Based on this interpretation and assuming $N$ is large, the analysis for the continuous UA can be straightforwardly extended to the case of discrete UA by including the quantization error. As a result, for channel estimation with single pilot symbols, the channel observation profile  for the discrete UA, denoted as $\hat{\Phi}(Y)$, can be written as 
\begin{equation}
\hat{\Phi}(Y) = \Phi(Y) + O\l(\frac{1}{N}\r). 
\end{equation}
where $\Phi(Y)$ for the continuous UA is given in Theorem~\ref{Theo:Profile}. Moreover, the result in Theorem~\ref{Theo:Profile:Seq} for the case of pilot sequences can be modified by replacing $o$ with $o + O(1/N)$. Similarly, for data transmission using the discrete UA,  the receive SINR at user $u$, denoted as $\widehat{\SINR}_u$, can be shown to be $\widehat{\SINR}_u = \SINR_u + O(1/N)$ with $\SINR_u$ corresponding to the continuous UA. 
The above results suggest  that with respect to the continuous counterpart, the discrete UA causes additional fluctuation in the channel observation profile  and receive SNRs, which can potentially degrades the performance of channel estimation and data transmission. 

\section{Communication Using the Circular UA: Data Transmission}\label{Section:DataTx}
In this section, two precoding  techniques are designed for the UA, namely the  channel conjugate and the \emph{multiuser phase mode} (MU-PM) precoding. For simplicity, it is assumed that the UA has perfect knowledge of the LI-channels.

\subsection{Channel Conjugate Transmission}
For channel conjugate transmission, the precoder applies a phase shift to each antenna for compensating propagation delay to achieve coherent  combining at the target user location, which is similar to beamforming using a phase array. Specifically, the precoder $f_u(\phi)$ is given as 
\begin{equation}
  f_u(\phi) = \frac{h_u^*(\phi)}{|h_u(\phi)|},\quad \phi \in [0, 2\pi)\label{Eq:CC:Prec}
\end{equation}
where $h(\phi)$ is given in \eqref{Eq:Ch:Gain}. It follows that 
\begin{equation}\label{Eq:CCTx:Prec}
f_u(\phi)  =  e^{j\frac{2\pi }{\lambda} (r_0 - r_u\psi_u(\phi))}, \quad \phi \in [0, 2\pi). 
\end{equation}
The normalization $|f_u(\phi)|^2 = 1$ facilitates the UA implementation under the per-element power constraint  e.g., using a  phase array.

The channel conjugate precoder $\{f_u(\phi)\}$   shapes the distribution of the field power density  such transmission power is concentrated in a small region centered at  the target   location $X_u$.  Characterizing the distribution is useful for analyzing the precoder performance. To this end, consider the transmission of an unmodulated wave using the UA  after precoding using $\{f_u(\phi)\}$ in \eqref{Eq:CCTx:Prec}.  Conditioned on the precoder, let $g(X_k\mid X_u)$  denote the resultant  field  measured at the location $X_k$. With the propagation distance in \eqref{Eq:Dist:Approx}, it can be obtained that 
\begin{equation}
g(X_k\mid X_u) =\sqrt{\frac{P_{\textrm{t}}}{2\pi r_0}}\times \frac{1}{\sqrt{4\pi r_0^2}}\times \int_0^{2\pi}  e^{j\frac{2\pi}{\lambda}(-r_0 + r_u\cos\psi_u(\phi) ) } f_k(\phi) r_0 d \phi + \frac{o}{\sqrt{r_0}}. \label{Eq:Field:Dist}
\end{equation}
At the right-hand side of \eqref{Eq:Field:Dist}, the three factors of the dominant term correspond to the density of  transmission power uniformly distributed over the UA, the propagation loss and the wave superposition at $X_u$, respectively. Substituting the precoder in \eqref{Eq:CCTx:Prec} into \eqref{Eq:Field:Dist} yields 
\begin{align}
g(X_k\mid X_u) &= \sqrt{\frac{P_{\textrm{t}}}{2 r_0}}\times \frac{1}{2\pi} \int_0^{2\pi}  e^{j\frac{2\pi}{\lambda}(r_u\cos\psi_u(\phi)- r_k\cos\psi_k(\phi)) } d \phi   + \frac{o}{\sqrt{r_0}}. \label{Eq:Field:Dist:cir}
\end{align}
The  field power density at location $X_k$ can be represented by  $p(X_k\mid X_u) = |g(X_k\mid X_u)|^2$.  Using \eqref{Eq:Field:Dist:cir}, a  closed-form expression for $p(X_k\mid X_u) $ is obtained as shown in the following lemma proved in Appendix~\ref{App:CC:Field}. 

\begin{lemma}[Field power density distribution] \label{Lem:CC:Field}
Given the circular UA and channel conjugate precoding targeting user $X_u$, the field power density  measured at  the user location $X_k$ is given as 
\emph{\begin{equation}
p(X_k\mid X_u) = \frac{P_{\textrm{t}}}{2 r_0}  J_0^2\l(\tfrac{2\pi}{\lambda}|X_u - X_k|\r)  + \frac{o}{r_0}.  \label{Eq:Field:Dist:cir:a}
\end{equation}}
\end{lemma}

\begin{figure}[t]
\begin{center}
\includegraphics[width=10cm]{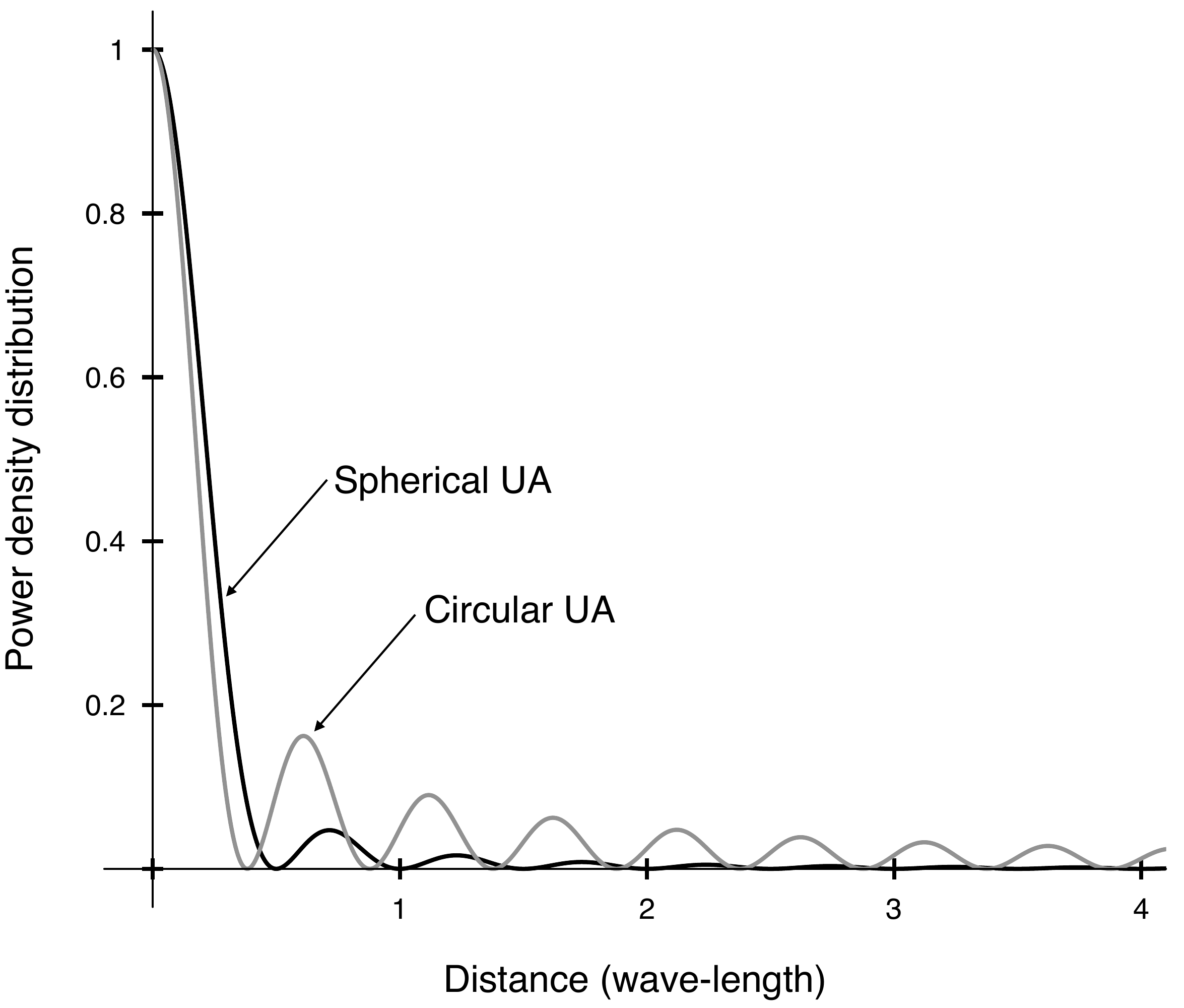} 
\caption{Given channel-conjugate transmission, the distribution of field power density (normalized by its peak) as a function of distance from the peak location.}
\label{Fig:Bessel:Bnd}
\end{center}
\end{figure}

The result shows that the field power density function  $p(X_k\mid X_u)$ depends only on the distance $|X_k - X_u|$ and thus can be rewritten as $p(d)$ with $d > 0$ denotes the the distance from the location targeted by the precoder. Then the distribution of the field power density can be characterized by the function $p(d)/p(0)  = J_0\l(\tfrac{2\pi d}{\lambda}\r)$ that is plotted in Fig.~\ref{Fig:Bessel:Bnd}. It can be observed from the figure that the channel-conjugate precoder shapes the field distribution such that most power is concentrated within a circular region centered at   the target location and having a radius of half wavelength. The tail of the distribution function is undesirable as it causes interference to nearby unintended receivers. However, the envelop of the tail decays with distance $d$, allowing interference suppression by spatial separation as further discussed in the sequel. 

With the field distribution in \eqref{Lem:CC:Field}, the performance of the channel-conjugate precoding is readily analyzed in terms of receive SINRs as follows.  Let $P_{\textrm{r}}$ and $P_{\textrm{i}}$ denote the signal and interference powers at user $X_u$, respectively. Since the effective aperture of the receive omni-directional antenna is $\lambda/4\pi$, $P_{\textrm{r}}$ and $P_{\textrm{i}}$ are given as 
\begin{equation}
P_{\textrm{r}} = \frac{\lambda}{4\pi }p(X_u\mid X_u) \quad\textrm{and}\quad P_{\textrm{i}} = \frac{\lambda}{4\pi } \sum_{k \neq u}p(X_u\mid X_k). \label{Eq:Pr:Pi} 
\end{equation}
The receive SINR for user $X_u$ can be written in terms $P_{\textrm{r}}$ and $P_{\textrm{i}}$ as 
\begin{equation}
\SINR_u = \frac{P_{\textrm{r}}}{P_{\textrm{i}} + \sigma^2}. \label{Eq:SINRu}
\end{equation}
Substituting  Lemma~\ref{Lem:CC:Field}, \eqref{Eq:Pr:Pi}  into \eqref{Eq:SINRu} yields the following main result of this section. 

\begin{theorem}[Receive SINRs] \label{Theo:ChanConj:Tx}
For  channel conjugate  transmission using the circular UA, the receive SINR for user  $u$ is given as
\begin{equation}
\SINR_{u}  = \frac{1}{\sum\limits_{k\neq u } J_0^2\left(\frac{2\pi}{\lambda}|X_{u} - X_k|\right) + \frac{1}{\SNR} }+o, \qquad u = 1, 2, \cdots, U \label{Eq:SINR:CC}
\end{equation}
where the receive SNR is given as 
\begin{equation}
\SNR = \frac{P_{\textrm{t}} \lambda}{8\pi \sigma^2 r_0}  + \frac{o}{r_0}. \label{Eq:SNR:Cir}
\end{equation}
\end{theorem}

\begin{remark}[High SNR]\label{Re:CC:SIR}\emph{
Applying the bound on the Bessel function in Property (B$4$) in Appendix~\ref{App:Bessel}, for a high SNR, the signal-to-interference ratio (SIR)  at user $u$ scales with the distance to the nearest interferer, namely $\min_{k\neq u}|X_k - X_u|$,  and the wavelength $\lambda$ as
\begin{equation}
\SIR_{u}  \geqslant \frac{\nu^2}{U-1}\l(2\pi\min_{k\neq u}\frac{|X_k - X_u|}{\lambda}\r)^{\frac{2}{3}} + o, \qquad \forall \ u.  \label{Eq:SIR}
\end{equation}
Thus the receive SIRs increase with the increasing minimum user-separation distance (in wavelength)  following a \emph{sub-linear} function. Moreover,  the result in \eqref{Eq:SIR} suggests that denser simultaneous users  can be supported by reducing the wavelength without compromising  the system throughput. Specifically, the user density can scale linearly with $1/\lambda^2$. 
}
\end{remark}

\begin{remark}[Free space vs. scattering channels]\emph{Free-space channels allow the UA to focus signal energy at intended users such that the signal power decays rapidly with the distance from the target location. As a result, given single-user (channel-conjugate) transmission, interference can be suppressed by increasing user spatial separation as shown in \eqref{Eq:SIR}. However, this is infeasible in the scenario of scattering channels  as scattering introduces additional cross coupling between multiuser signals. Suppressing the resultant interference cannot rely on increasing  users' spatial separation and multiuser precoding has  to be used for this purpose. }
\end{remark}

\begin{remark}[High Mobility]\emph{Given fixed user locations, the sum rate   is  thus $\sum_u \log_2(1 +\SINR_u)$ with $\SINR_u$ given in \eqref{Eq:SINR:CC}. For the scenario where users have high mobility, it is more appropriate to consider the ergodic sum rate given as 
\begin{equation}
\bar{R} = \E\l[\sum_u \log_2(1 +\SINR_u(X_1, \cdots, X_U))\r]
\end{equation}
where $\{X_u\}$ are random. The ergodic sum rate can be analyzed by combining the results in Theorem~\ref{Theo:ChanConj:Tx} and stochastic geometry \cite{StoyanBook:StochasticGeometry:95}. 
}
\end{remark}

Last, the received SNR given in \eqref{Eq:SNR:Cir} implies the following result. 
\begin{corollary}[Propagation loss] \label{Cor:PathLoss}Given channel-conjugate transmission using the circular UA,  the propagation loss is given as \emph{
\begin{equation}
\frac{P_{\textrm{r}}}{P_{\textrm{t}}} = \frac{\lambda}{8\pi r_0}+ \frac{o}{r_0}. 
\end{equation}}
\end{corollary}
In other words, transmission using the circular UA reduces the loss such that it is inversely  proportional to the propagation distance instead of its square as in the case of a conventional array of collocated antennas.

\subsection{Multiuser Phase Mode  Precoding}
In this section, MU-PM precoders are designed under the zero-forcing constraints to avoid  multiuser interference.  Exploiting the circular structure of the UA, the precoder for each user, say $  f_u'(\phi)$ for user $u$, can be expressed in terms of  the  Fourier series representing the sequence of phase modes:  
\begin{equation}\label{Eq:Precod:ZF}
  f_u'(\phi) =   f_u(\phi) \sum_{m=-\infty}^\infty c_{u, m} e^{-jm\phi}, \qquad \phi \in [0, 2\pi)
\end{equation}
where $f_u(\phi) $ is the channel-conjugate precoder in \eqref{Eq:CC:Prec} for compensating the channel phase shift and the summation is the said Fourier series.   The precoder   coefficients  $\{c_{u, m}\}$ satisfy   the power   constraint:  $\sum_{m=-\infty}^\infty |c_{u, m}|^2 \leq 1$ for all $u$.

The precoder coefficients are designed to avoid  the inter-user interference as follows. Without multiuser interference, it is unnecessary to analyze the field spatial distribution as in the case of channel conjugate transmission and instead the remainder of the section focuses on the precoder design and the analysis of the received signal power.  To this end, the received signal at user $u$ is obtained  by substituting the precoder in  \eqref{Eq:Precod:ZF} into \eqref{Eq:Sig:DL}: 
\begin{equation}
 y_{u}  = \sqrt{\frac{\lambda P_t}{8\pi r_0}} \sum_{k=1}^U \l( \frac{1}{2\pi} \int_0^{2\pi} \frac{h^*_k(\phi)h_u(\phi)}{|h_{k}(\phi)|^2} \sum_{m=-\infty}^\infty c_{k, m}   e^{-jm\phi}\ d\phi\r)  x_{k} + z_{u}.  \nn
\end{equation}
The substitution of the channel response in \eqref{Eq:Ch:Gain} yields  
\begin{equation}
 y_{u} = \sqrt{\frac{\lambda P_t}{8\pi r_0}} c_{u, 0} x_u  + \sqrt{\frac{\lambda P_t}{8\pi r_0}} \sum_{k\neq u}\sum_{m=0}^\infty c_{k, m}  \mathcal{J}_{u, k, m}   x_{k}+z_{u} +\frac{o}{r_0}\label{Eq:Sig:ZF}
\end{equation}
where the first term and the summation represent the signal and  interference, respectively, and $\mathcal{J}_{u, k, m}$ is defined as   
\begin{equation}\label{Eq:J:Def}
\mathcal{J}_{u, k, m} = \frac{1}{2\pi } \int_0^{2\pi} e^{j\frac{2\pi}{\lambda}\left(r_{{u}}\cos(\varphi_{{u}}-\phi)- r_{{k}}\cos(\varphi_{{k}}-\phi)\right)-jm\phi} \ d\phi.  
\end{equation}
A closed-form expression for $\mathcal{J}_{u, k, m}$ can be obtained as follows,  following similar steps as in the proof for Lemma~\ref{Lem:CC:Field} with the details omitted for brevity.  
\begin{lemma} \label{Lem:J:Coeff}The coefficients $\{\mathcal{J}_{u, k, m}\}$ defined in \eqref{Eq:J:Def} can be written as 
\begin{equation}
\mathcal{J}_{u, k, m} = e^{jm \beta_{u, k}}J_m\left(\frac{2\pi}{\lambda}|X_u - X_k|)\right) \label{Eq:CrossTerm} 
\end{equation}
where $1 \leq u, k \leq U$ and $m$ is an integer. 
\end{lemma}
The interference term in the signal in \eqref{Eq:Sig:ZF} can be nulled by enforcing the following zero-forcing constraints: 
\begin{equation}\label{Eq:ZF}
  \sum_{m=-\infty}^\infty c_{k, m} \mathcal{J}_{u, k, m} = 0 , \quad \forall \ k \neq u.
\end{equation}
In practice, given a discrete UA and a constraint on computation complexity, it is infeasible to implement MU-PM precoders with an infinite number of phase modes and only a finite set of modes is considered in the design. Let the corresponding  coefficient set be denoted as  $\{c_{u, m} \mid1\leq  u \leq U, -M \leq m \leq M\}$ with $M$ being a fixed integer and other coefficients set to zero. Under the zero-forcing constraints in \eqref{Eq:ZF},  it is desirable to choose the precoder coefficients such that the  correspond set of coefficients $\{\mathcal{J}_{u, k, m}\mid 1 \leq u, k\leq U, -M \leq m \leq M\}$ are significant. Then applying Property (B4) of Bessel functions  in Appendix~\ref{App:Bessel} gives that 
\begin{equation}\label{Eq:DoF}
M =2\pi \min\limits_{u, k}\left\lfloor\frac{ |X_u - X_k|}{\lambda}\right\rfloor 
\end{equation}
where users are assumed to be separated by distances much larger than a single wavelength,  yielding  the following fact. 

\begin{remark}[Degrees of freedom] \emph{
Given MU-PM transmission using a circular UA, the total number of degrees of freedom for interference avoidance is approximately equal to $(2M+1)$ with $M$ given in \eqref{Eq:DoF}, which is  proportional to the minimum user-separation distance.  
}\end{remark}


Transmission over only $(2M+1)$ phase modes allows  the precoder coefficients for user $u$ to be represented by the vector $\bc_{u} = \begin{bmatrix}c_{u, -M}, &\cdots,  & c_{u, 0},  & \cdots, & c_{u, M} \end{bmatrix}^T$. Furthermore,  the zero-forcing constraints in \eqref{Eq:ZF} can  be written in a matrix form: 
\begin{equation}\label{Eq:Approx:ZF}
   \mathbfcal{J}_{u}\bc_{u} =\b0
\end{equation}
where the matrix $\mathbfcal{J}_{u}$ is defined as 
\begin{equation}\label{Eq:ZF:Coeff}
  \mathbfcal{J}_{u} = \begin{bmatrix} \cJ_{u, 1, -M}, &\cdots, &  \cJ_{u, 1, 0}, & \cdots, &  \cJ_{u, 1, M} \\
                                       \vdots & \ddots &  \vdots   &\ddots& \vdots  \\
                                       \cJ_{u, u-1, -M}, &\cdots, &\cJ_{u, u-1, 0}, & \cdots, &  \cJ_{u, u-1, M} \\
                                       \cJ_{u, u+1, -M}, &\cdots, &   \cJ_{u, u+1, 0}, & \cdots, &  \cJ_{u, u+1, M} \\
                                       \vdots & \ddots & \vdots   &\ddots & \vdots  \\
                                       \cJ_{u, U, -M}, &\cdots,   &   \cJ_{u, U, 0},  &\cdots, &  \cJ_{u, U, M}
                                        \end{bmatrix}
\end{equation}
with the elements specified in Lemma~\ref{Lem:J:Coeff}.
Given the  constraints  obtained in \eqref{Eq:Approx:ZF}, the main result of the section can be readily stated as follows. 
 
\begin{theorem}[Multiuser UA precoding] \label{theorem:SZF_solutions}
For MU-PM transmission using the circular UA,  the   precoder coefficients are given as 
\emph{
\begin{equation}\label{Eq:ZF:Solution}
  \bc_{u} \in \textsf{null}(\mathbfcal{J}_{u}), \qquad u = 1, 2, \cdots,  U
\end{equation}
}
where the matrix $\mathbfcal{J}_{u}$ is given in \eqref{Eq:ZF:Coeff} and  \emph{$\textsf{null}(\mathbfcal{J}_{u}) $} denotes the null-space of $\mathbfcal{J}_{u}$.  
\end{theorem}

With interference avoided, the receive SNR for user $u$ follows from \eqref{Eq:Sig:ZF} as 
\begin{equation}
\SNR_u = \frac{\lambda P_t |c_{u, 0}|^2}{8 \pi r_0 \sigma^2 }+\frac{o}{r_0}=\frac{\lambda P_t |\bee_0^\dag\bc_{u}|^2}{8 \pi r_0 \sigma^2 } + \frac{o}{r_0}
\end{equation}
where $\bee_0 = [0,\cdots,0,1,0,\cdots,0]^T$. Thus, to maximize $\SNR_u$, the precoder-coefficient vector $\bc_{u}$ should be chosen as the projection of $\bee_0$ onto $\textsf{null}(\mathbfcal{J}_{u})$ that contains $\bc_{u}$ according to Theorem~\ref{theorem:SZF_solutions}. This gives the following corollary. 

\begin{corollary}\label{Cor:RxSNR}\emph{For MU-PM transmission using the circular UA, the maximum receive SNRs are given as 
\begin{equation}\label{Eq:SNR:ZF}
\SNR_u =\frac{\lambda P_t \left|\bee_0^\dagger \bb_{u}\right|^2}{8 \pi r_0 \sigma^2 }  + \frac{o}{r_0}, \qquad u = 1, 2, \cdots, U
\end{equation}
where $\bb_{u}$ is a basis of $\textsf{null}(\mathbfcal{J}_{u})$. 
}
\end{corollary}

With respect to the case of channel-conjugate transmission, the term $\left|\bee_0^\dagger \bb_{u}\right|^2$ represents the loss in receive SNR due to interference nulling.

\section{Communication Using the  Spherical UA}\label{Section:Spherical:UA}

In the preceding section, the UA is modeled as the circular array. The results are extended  in this section to the spherical UA system, which shows its performance improvements with respect to the circular-UA counterpart. Essentially, the difference in analysis arises from the use of spherical harmonics as the tool in place of Fourier series.

\subsection{Spherical  LI-Channel Estimation}\label{Section:ChanEst:Sph}
Assume that users transmit single pilot symbols: $\{s_u\} = \{1\}$. The estimation scheme in Section~\ref{Section:ChanEst} is extended to the spherical UA as follows. To this end,  the training signal in \eqref{Eq:Training:Sig}, re-denoted  as $q(\phi, \theta)$,  is expanded as a Laplace series as follows by  using the harmonic functions $\{Y^m_{\ell}\}$,  defined in  \eqref{Eq:Sph:Basis} in Appendix~\ref{App:Bessel}, as the basis: 
\begin{equation}\label{Eq:Train:Sph}
q(\phi, \theta) = \sum_{\ell = 0}^\infty \sum_{m = -\ell}^\ell Q_{\ell}^m Y^m_{\ell}(\phi, \theta), \qquad \theta \in [0, \pi), \phi \in [0, 2\pi) 
\end{equation}
where the Laplace coefficients $\{Q_{\ell}^m\}$ are defined as 
\begin{equation}\label{Eq:Q:Sph}
Q_{\ell}^m = \sum_{u=1}^U\int_{\theta=0}^{2\pi} \int_{\psi=0}^\pi q(\phi, \theta) Y^m_{\ell}(\phi, \theta)  \sin\phi d\phi d\theta. 
\end{equation}
To derive a closed-form expression for $Q_{\ell}^m$, a useful result is given as follows. 


\begin{lemma}\label{Lem:Sph:Expan} Given two points $(r_0, \phi, \theta), (r_u, \phi_u, \theta_u)\in \mathds{R}^3$, 
\begin{equation}\label{equ:3d_first_term_expand}
  e^{j\frac{2\pi}{\lambda} r_u\cos \psi_u } = (2\pi)^{\frac{3}{2}} \sum_{\ell=0}^\infty \sum_{m=-\ell}^\ell \frac{j^\ell J_{\ell+\frac{1}{2}}\left(\frac{2\pi}{\lambda}r_u\right)}{\left(\frac{2\pi}{\lambda}r_u\right)^{\frac{1}{2}}}  Y^m_{\ell}(\phi_u, \theta_u)  Y^m_{\ell}(\phi, \theta)
\end{equation}
where $\psi_u$ denotes the angle between the vectors corresponding to  the two points. 
\end{lemma}
\noindent The lemma is proved in Appendix~\ref{Lem:Sph:Expan:Proof}.  Using  the orthogonality of  the basis functions $\{Y^m_{\ell}(\phi, \theta)\}$, substitution of the expression for $q(\phi, \theta)$ in \eqref{Eq:Training:Sig} and Lemma~\ref{Lem:Sph:Expan} into \eqref{Eq:Q:Sph} leads to the following lemma. 

\begin{lemma}[Training signal decomposition]\label{Lem:Qmn}The training signal received at the spherical UA, namely $\{q(\phi, \theta)\}$ in \eqref{Eq:Train:Sph}, has the Laplace  coefficients given as follows: 
\begin{equation}\label{Eq:Qml}
Q_{\ell}^m  = \frac{\lambda (2\pi)^{\frac{3}{2}}}{4\pi r_0}  \sum_{u=1}^U \frac{j^\ell J_{\ell+\frac{1}{2}}\left(2\pi r_u/\lambda\right)}{\left(2\pi r_u/\lambda\right)^{\frac{1}{2}}}  Y^m_{\ell}(\phi_u, \theta_u)  + \frac{o}{r_0}
\end{equation}
where $\ell = 1, 2, \cdots$ and $-\ell \leq m \leq \ell$.
\end{lemma}

\noindent Define two sets  $\tilde{\bQ}$ and $\tilde{\bV}(Y)$ as $\tilde{\bQ} = \{Q_{\ell}^m\}$ and $\tilde{\bV}(Y) = \{V_{\ell}^m(Y)\}$ with their elements $Q_{\ell}^m$ given in \eqref{Eq:Qml} and 
  $V_{\ell}^m: \mathds{R}^3\rightarrow \mathds{R}$ being a function defined as 
\begin{equation}\label{Eq:Vml}
V_{\ell}^m(Y) = (2\pi)^{\frac{3}{2}}  \frac{j^n J_{\ell+\frac{1}{2}}\left(\frac{2\pi}{\lambda}r_u\right)}{\left(\frac{2\pi}{\lambda}r_u\right)^{\frac{1}{2}}}  Y^m_{\ell}(\phi_u, \theta_u).  
\end{equation}
Using these definitions, the channel observation profile  corresponding to the spherical UA, represented by $\tilde{\Phi}(Y)$,   can be defined similarly as in  \eqref{Eq:Loc:Profile}: 
\begin{align}
\tilde{\Phi}(Y) &=\frac{r_0}{\lambda }\left | \tilde{\bV}(Y) \circ\tilde{\bQ}  \right |\nn\\
&=\frac{r_0}{\lambda }\left |\sum_{u=1}^U \sum_{\ell=0}^\infty\sum_{m=-\ell}^\ell  [V_{\ell}^m(Y)]^*Q_{\ell}^m\right |.\nn  
\end{align}
Substituting \eqref{Eq:Qml} and \eqref{Eq:Vml} gives 
\begin{equation}
\tilde{\Phi}(Y) =   \left|2\pi^2 \sum_{u=1}^U \sum_{\ell=0}^\infty \frac{J_{\ell+\frac{1}{2}}\left(\frac{2\pi r_u}{\lambda}\right)J_{\ell+\frac{1}{2}}\left(\frac{2\pi r_Y}{\lambda}\right)}{\left(\frac{ 2\pi r_u}{\lambda}\right)^{\frac{1}{2}}\left(\frac{ 2\pi r_Y}{\lambda}\right)^{\frac{1}{2}}} \sum_{m=-\ell}^\ell [Y^m_{\ell}(Y)]^* Y^m_{\ell}(X_u) \right| +o
\label{Eq:Prof:Sph}
\end{equation}
where noise varnishes for the same reason as for the case of circular UA. 
By applying  Addition Theorem in Property  (S3) for spherical harmonics in Appendix~\ref{App:Bessel}, 
\begin{equation}
\tilde{\Phi}(Y) = \left|\frac{\pi}{2}  \sum_{u=1}^U \sum_{\ell=0}^\infty \frac{(2\ell+1)J_{\ell+\frac{1}{2}}\left(\frac{2\pi r_u}{\lambda}\right)J_{\ell+\frac{1}{2}}\left(\frac{2\pi r_Y}{\lambda}\right)}{\left(\frac{ 2\pi r_u}{\lambda}\right)^{\frac{1}{2}}\left(\frac{ 2\pi r_Y}{\lambda}\right)^{\frac{1}{2}}}P_\ell\left(\cos\psi_u\right) \right|^2 +o. 
\end{equation}
Next, applying Addition Theorem in Property (B3) of  Bessel functions in Appendix~\ref{App:Bessel} further simplifies the expression  as shown in the following theorem. 

\begin{theorem}[Channel observation] \label{Theo:Profile:Sph} The channel observation profile  corresponding to the spherical UA is given as 
\begin{equation}
\tilde{\Phi}(Y) = \left|\sum_{u=1}^U\mathrm{sinc}\left(\tfrac{ 2\pi}{\lambda}|Y - X_u|\right) \right|   + o, \qquad \textrm{a.s.}  
\end{equation}
\end{theorem}

\begin{remark}[Channel estimation error] \emph{As in Remark~\ref{Re:EstErr} for the  circular UA, the accuracy of channel estimation can be evaluated using the difference $  \l|\tilde{\Phi}(X_u) - 1 \r|$.  Since  $\left|\mathrm{sinc}(d) \right| \leqslant d^{-1}$, it can be obtained from Theorem~\ref{Theo:Profile:Sph} that 
\begin{align}
  \l|\tilde{\Phi}(X_u) - 1 \r| 
  & \leq 2\pi (U - 1) \min_{u \neq k} \l(\tfrac{|X_k - X_u|}{\lambda} \r)^{-1} + o, \quad \text{a.s.} \label{Eq:ErrBnd:Sph}
\end{align}
The error bound  is observed to diminish inversely with the  increasing minimum user-separation distance (in wavelength) following an inverse function that is faster than the circular-UA counterpart in \eqref{Eq:ErrBnd}. This quantifies the gain of increasing the UA by one dimension from the perspective  of channel estimation. 
}
\end{remark}

Next, consider the case where users transmit pilot sequences with length of $L$ symbols.  Following the same procedure as for deriving Theorem~\ref{Theo:Profile:Seq} yields the following corollary. 

\begin{corollary}[Effect of pilot sequences]\label{Theo:Profile:Seq:Sph}
Given $L \geq U$ and orthogonal pilot sequences, the LI-channel estimation using the spherical   UA is almost perfect since the channel observation profile  is approximately equal to the single-user counterpart:
\emph{
\begin{equation}
\Big |\tilde{\Phi}_u(Y) -  \mathrm{sinc}\left(\tfrac{2\pi}{\lambda}|X_{u} - Y|\right) \Big |\leqslant  o. \nn
\end{equation}
}
\end{corollary}
\begin{remark}\emph{If orthogonal pilot sequences are used, the spherical UA does not have an advantage over the circular counterpart in terms of channel estimation. However, the former improves the performance of channel estimation in the case of single pilot symbols as well as that of  data communication  as shown in the sequel. 
}
\end{remark}

\subsection{Data Transmission Using the Spherical UA}
\subsubsection{Channel Conjugate Transmission}
For channel conjugate transmission using the spherical UA, the spherical precoder, denoted as  $\tilde{f}_u(\phi, \theta)$,  is modified from the circular counterpart in \eqref{Eq:CCTx:Prec} as  
\begin{equation}\label{Eq:CCTX:Sph}
\tilde{f}_u(\phi, \theta) = \frac{h_u^*(\phi, \theta)}{|h_u(\phi, \theta)|},\quad  \theta\in [0,  2\pi), \phi \in [0,   \pi).  
\end{equation}

The resultant receive SNRs are derived as follows.  Let $\tilde{q}(X_u \mid X_k)$ and $\tilde{p}(X_u \mid X_k)$ denote the field and its power density  measured at location $X_u$ given a precoder targeting user $X_k$. Then  $\tilde{q}(X_u \mid X_k)$ can be obtained by modifying the circular-UA counterpart   in \eqref{Eq:Field:Dist} by replacing the integration over a circle with one over a sphere: 
\begin{align}
\tilde{g}(X_u \mid X_k) &=\sqrt{\frac{P_{\textrm{t}}}{4\pi r_0^2}}\times \frac{1}{\sqrt{4\pi r_0^2}}\times \iint e^{j\frac{2\pi}{\lambda}(r_{u}\cos\psi_{u}- r_{k}\cos\psi_k) } r_0^2 \sin \theta \ d \theta \ d \phi + o \nn\\
&= \frac{\sqrt{P_{\textrm{t}}}}{4\pi } \iint e^{j\frac{2\pi}{\lambda}(r_{u}\cos\psi_{u}- r_{k}\cos\psi_k) } \sin \theta \ d \theta \ d \phi + o. 
\label{Eq:Field:Dist:Sph}
\end{align}
To facilitate analysis, a set of coefficients $\{\mathcal{J}_{u, k, m, n}\}$ with integer indices $(u, k, m, n)$ are defined as 
\begin{equation}
\mathcal{J}_{u, k, m, \ell} =  \iint e^{j\tfrac{2\pi}{\lambda} (r_u\cos\psi_u - r_k\cos\psi_k) } Y^m_{\ell}(\phi, \theta) \sin\phi \ d\phi \ d\theta    \label{Eq:J:Sph:Def}
\end{equation}
with $1 \leq u, k \leq U$, $n \geq 0$ and $-\ell \leq m \leq \ell$, which are  also used for designing multiuser precoders in the next sub-section. They can be written in a closed form as shown in the following lemma that is proved in Appendix~\ref{App:J:Spy:Proof}. 
 
 \begin{lemma}\label{Lem:J:Spy} The coefficient $\mathcal{J}_{u, k, m, \ell}$ defined in \eqref{Eq:J:Sph:Def} can be written as 
\begin{equation}
\mathcal{J}_{u, k, m, \ell} =  \frac{(2\pi)^{\frac{3}{2}}j^\ell J_{\ell+\frac{1}{2}}\left(\tfrac{2\pi}{\lambda} | X_u - X_k|\right)}{\sqrt{\tfrac{2\pi}{\lambda} | X_u - X_k|}}  Y^m_{\ell}(\phi_{u, k}, \theta_{u, k})  \nn
\end{equation}
where the angles $\phi_{u, k}$ and $\theta_{u, k}$ are defined by the following equations: 
\begin{align}
\sin\phi_{u, k}\cos\theta_{u, k} & = \frac{r_u\sin\phi_u\cos\theta_u - r_k\sin\phi_k\cos\theta_k}{r_{u, k}}\nn\\
\sin\phi_{u, k}\sin\theta_{u, k} &= \frac{r_u \sin\phi_u\sin\theta_u - r_k \sin\phi_k\sin\theta_k}{r_{u, k}}\nn\\
\cos \phi_{u, k} &= \frac{r_u\cos \phi_u - r_k\cos \phi_k}{r_{u, k}}. 
\end{align}
\end{lemma}

Using the fact that 
\begin{equation}
J_{\frac{1}{2}}(x)  = \sqrt{\frac{2}{\pi}} \frac{\sin x}{\sqrt{x}}, \qquad Y_{0, 0}(\phi_{u, k}, \theta_{u, k}) = \frac{1}{\sqrt{4\pi}}
\end{equation}
and Lemma~\ref{Lem:J:Spy},   the  field in \eqref{Eq:Field:Dist:Sph} can be obtained in a closed form, yield the following result. 

\begin{lemma}[Field power density distribution] \label{Lem:CC:Field:Sph}
Given the spherical  UA and channel conjugate precoding targeting user $X_u$, the field power density  measured at  the user location $X_k$ is given as 
\begin{equation}
\tilde{p}(X_u \mid X_k) = P_{\textrm{t}}\ \mathrm{sinc}^2\l(\tfrac{2\pi}{\lambda}|X_u - X_k|\r)  + o. \label{Eq:Field:Dist:a}
\end{equation}
\end{lemma}

As in the case of  circular UA, $\tilde{p}(X_u \mid X_k)$ can be rewritten as $\tilde{p}(d)$ with $d$ being the distance from the location targeted by the precoder. Then  the distribution of the field power density can be characterized by the function $\tilde{p}(d)/\tilde{p}(0) = \mathrm{sinc}^2\l(\tfrac{2\pi d}{\lambda}\r)$ that is plotted in Fig.~\ref{Fig:Bessel:Bnd}. Like the circular-UA counterpart, by  channel-conjugate precoding, the spherical UA focuses  the transmission power into  a  region within a distance of  half wavelength from    the target location. The advantage  of the spherical UA  is reflected in  that the tail of the distribution function has an envelop diminishing with the increasing distance much faster than that corresponding to the circular UA. This reduces  multiuser interference and leads to substantial  performance improvements as shown in the analysis and observed from simulation results. 

Given the SINR defined similarly as in \eqref{Eq:SINRu}, the  result in Lemma~\ref{Lem:CC:Field:Sph} leads to  the spherical-UA counterpart of Theorem~\ref{Theo:ChanConj:Tx} as follows. 

\begin{theorem}[Receive SINRs] \label{Theo:ChanConj:Sph}
For  channel conjugate  transmission using the spherical UA, the receive SINR for user  $u$, denoted as $\widetilde{\SINR}$  is given as
\begin{equation}
\widetilde{\SINR}_{u}  = \frac{1}{\sum\limits_{k\neq u } \mathrm{sinc}^2\left(\frac{2\pi}{\lambda}|X_{u} - X_k|\right) + \frac{1}{\SNR} }+o, \qquad u = 1, 2, \cdots, U \label{Eq:SINR:CC:Sph}
\end{equation}
where the receive SNR is 
\begin{equation}\label{Eq:RxSNR:Sph}
\SNR = \frac{P_{\textrm{t}} \lambda^2}{4\pi \sigma^2}. 
\end{equation}
\end{theorem}

\begin{remark}[High SNR] \emph{Using the fact $\left|\mathrm{sinc}(d) \right| \leqslant d^{-1}$, the SIR lower bound in \eqref{Eq:SIR} for the circular UA can be modified for the current case as 
\begin{equation}
\widetilde{\SIR}_{u}  \geqslant \frac{1}{U-1}\l(2\pi\min_{k\neq u}\frac{|X_k - X_u|}{\lambda}\r)^2 + o.  
\end{equation}
Thus, the receive SIRs increase with the minimum user-separation distance (in wavelength) at least following a super-linear function  with the exponent $2$, which is faster than the sub-linear function for the circular UA (see Remark~\ref{Re:CC:SIR}).  This specifies the gain of increasing the UA by one dimension from the perspective of received signal quality. 
}
\end{remark}

Last, the received SNR in \eqref{Eq:RxSNR:Sph} suggests the following result. 

\begin{corollary}[Propagation loss] \label{Cor:PathLoss:sph} Given channel-conjugate transmission using the spherical  UA,  the propagation loss is approximately equal to the receive antenna aperture:  \emph{
\begin{equation}
\frac{P_{\textrm{r}}}{P_{\textrm{t}}} = \frac{\lambda^2}{4\pi} + o. 
\end{equation}}
\end{corollary}
In other words, the path loss is a constant and independent with the propagation distance $r_0$. In contrast, the loss corresponding to  the circular UA  and the conventional array is inversely proportional to $r_0$ (see Corollary~\ref{Cor:PathLoss}) and $r_0^2$, respectively.

\subsubsection{Multiuser Phase Mode Transmission}
The phase modes for the spherical UA correspond to different spherical harmonics. Then the spherical  MU-PM precoder for user $u$, denoted as  $  \tilde{f}_u'$, is  modified from the circular counterpart in \eqref{Eq:Precod:ZF} as
\begin{equation}\label{Eq:MuPrecod:Sph}
  \tilde{f}_u'(\phi, \theta) = \tilde{f}_u(\phi, \theta) \sum_{\ell=0}^\infty\sum_{m=-\ell}^\ell c_{u, m, \ell} Y^m_{\ell}(\phi, \theta)
\end{equation}
where $\{c_{u, m, \ell}\}$ are the precoder coefficients to be designed,   $\tilde{f}_u(\phi, \theta)$ is the channel conjugate precoder in \eqref{Eq:CCTX:Sph} and the spherical harmonic function $Y^m_{\ell}(\phi, \theta)$ is defined in \eqref{Eq:Sph:Basis}. By substitution of \eqref{Eq:MuPrecod:Sph} into \eqref{Eq:Sig:DL}, the signal received at user $X_u$ is given as 
\begin{align}
 \tilde{y}_{u} & =   \lambda\sqrt{\frac{P_t}{4\pi}} \sum_{k=1}^U \l( \frac{1}{4\pi } \iint \frac{h^*_k(\phi, \theta)h_u(\phi, \theta)}{|h_{k}(\phi, \theta)|}  \sum_{\ell=0}^\infty\sum_{m=0}^\ell c_{k, m, \ell} Y^m_{\ell}(\phi, \theta) \sin \phi d\phi d\theta  \r)  x_{k} + z_{u}   \nn\\
& =   \lambda\sqrt{\frac{P_t}{4\pi}} c_{u, 0, 0} x_u +  \lambda\sqrt{\frac{P_t}{4\pi}} \sum_{k\neq u }\l( \frac{1}{4\pi  } \iint e^{j\tfrac{2\pi}{\lambda} (r_u\cos\psi_u - r_k\cos\psi_k) }  \sum_{\ell=0}^\infty \sum_{m=-\ell}^\ell c_{k, m, \ell} Y^m_{\ell}(\phi, \theta) \sin \phi d\phi d\theta  \r)  x_{k} \nn\\
&\qquad + z_{u} +   o \nn\\   
&= \lambda\sqrt{\frac{P_t}{4\pi}} c_{u, 0, 0} x_u +   \lambda\sqrt{\frac{P_t}{4\pi}}  \sum_{k\neq u}\sum_{\ell=0}^\infty \sum_{m=0}^\ell c_{k, m, \ell}  \mathcal{J}_{u, k, m, \ell}   x_{k}+z_{u} + o\label{Eq:Sig:ZF:Sph}
\end{align}
where the coefficients $\{\mathcal{J}_{u, k, m, \ell}\}$ are defined earlier in \eqref{Eq:J:Sph:Def}.  It can be observed from Lemma~\ref{Lem:J:Spy} that $\mathcal{J}_{u, k, m, \ell}$ is proportional to $J_{\ell + \frac{1}{2}}(\tfrac{2\pi}{\lambda}|X_u - X_k|)$. Therefore, following the same reason as for the circular UA, only the set of coefficients $\{\mathcal{J}_{u, k, m, \ell}\mid  0 \leq \ell \leq M \}$ have significant values. Considering only these values reduces  the precoder coefficients to a finite set $\{c_{k, m, \ell}\mid  0\leq \ell \leq M\}$, yielding the following fact. 

\begin{remark}[Degrees of freedom]\label{Re:DoF:Sph}\emph{
Given MU-PM transmission using a spherical  UA, the total number of degrees of freedom for interference avoidance is approximately equal to $(M+1)^2$ with $M$ given in \eqref{Eq:DoF}. As a result, since $M$ is much greater  than one, the number of degrees of freedom generated by the spherical UA is approximately proportional to $M^2$ that is much larger than that, namely $2M$,  for the circular counterpart. 
}
\end{remark}

Next, the said finite set of precoder coefficients can be  designed by applying the zero-forcing constraints that follow from   \eqref{Eq:Sig:ZF:Sph} as
\begin{equation}\label{Eq:ZF:Sph}
  \sum_{\ell = 0}^M\sum_{m=-\ell}^\ell  c_{k, m, \ell} \mathcal{J}_{u, k, m, \ell} = 0 , \quad \forall \ k \neq u.
\end{equation}
The constraints can be written in the matrix form. To this end,  define the matrix $\mathbfcal{H}_{u, \ell}$  as 
\begin{equation}\label{Eq:ZF:Coeff:Sph}
  \mathbfcal{H}_{u, \ell} = \begin{bmatrix} \cJ_{u, 1, -\ell, \ell},  & \cJ_{u, 1, -\ell+1, \ell}, & \cdots &  \cJ_{u, 1, \ell, \ell} \\
                                       \vdots & \vdots    &\ddots& \vdots  \\
                                       \cJ_{u, u-1, -\ell, \ell},   &\cJ_{u, u-1, -\ell+1, \ell}, & \cdots &  \cJ_{u, u-1, \ell, \ell} \\
                                       \cJ_{u, u+1, -\ell, \ell},  &   \cJ_{u, u+1, -\ell+1, \ell}, & \cdots &  \cJ_{u, u+1, \ell, \ell} \\
                                       \vdots & \vdots    &\ddots & \vdots  \\
                                       \cJ_{u, U, -\ell, \ell},    & \cJ_{u, U, -\ell+1, \ell},  &\cdots &  \cJ_{u, U, \ell, \ell}
                                        \end{bmatrix}\nn 
\end{equation}
 and the row vector $\ba_{u, \ell} = \begin{bmatrix}c_{u, -\ell, \ell}, & c_{u, -\ell+1, \ell}, & \cdots& c_{u, \ell, \ell} \end{bmatrix}$ where $\ell = 0, 1, \cdots $ . Moreover, using these matrices/vectors as elements,   define 
\begin{equation}
\tilde{\mathbfcal{J}}_u =   \begin{bmatrix}\mathbfcal{H}_{u, 0}, \mathbfcal{H}_{u, 2}, \cdots, \mathbfcal{H}_{u, M}\end{bmatrix}, \qquad 
\bc_u =   \begin{bmatrix} \ba_{u, 1}, \ba_{u, 2}, \cdots, \ba_{u, \widetilde{M}}\end{bmatrix}^T. 
\end{equation}
Using these definitions, the zero forcing constraints in \eqref{Eq:ZF:Sph} can be written as 
\begin{equation}\label{Eq:Approx:ZF:Sph}
\tilde{\mathbfcal{J}}_u \tilde{\bc}_u =\b0, \qquad u = 1, 2, \cdots, U. 
\end{equation}

The main result of the section is summarized in the following theorem. 
\begin{theorem}[Multiuser UA precoding] \label{theorem:MPM:Tx:SPh}
For MU-PM transmission using the circular UA,  the   precoder coefficients under the zero-forcing constraints are given as 
\emph{
\begin{equation}\label{Eq:ZF:Solution:Sph}
 \tilde{\bc}_{u} \in \textsf{null}(\tilde{\mathbfcal{J}}_{u}), \qquad u = 1, 2, \cdots,  U
\end{equation}}
where $\textsf{null}(\tilde{\mathbfcal{J}}_{u}) $ denotes the null-space of $\tilde{\mathbfcal{J}}_{u}$.  
\end{theorem}
The spherical-UA counterpart of  Corollary~\ref{Cor:RxSNR} is as follows.  

\begin{corollary}[Receive SNRs]\label{Cor:RxSNR:Sph}For MU-PM transmission using the circular UA, the maximum receive SNR for user $u$ is given as 
\emph{
\begin{equation}\label{Eq:SNR:ZF:Sph}
\SNR_u = \frac{\eta^2 P_{\textrm{t}} \left|\bee_0^\dagger \bb_{u}\right|^2}{\sigma^2} + o, \qquad u = 1, 2, \cdots, U
\end{equation}}
where $\bb_{u}$ is a basis of \emph{$\textsf{null}(\tilde{\mathbfcal{J}}_{u})$}. 
\end{corollary}

\begin{remark}\emph{Theorem~\ref{theorem:MPM:Tx:SPh} and Corollary~\ref{Cor:RxSNR:Sph} are observed to have the same forms as Theorem~\ref{Theo:ChanConj:Tx} and Corollary~\ref{Cor:RxSNR}, respectively. However, the space $\textsf{null}(\tilde{\mathbfcal{J}}_{u})$ corresponding to the sperical UA is much larger than its circular-UA counterpart $\textsf{null}(\mathbfcal{J}_{u})$. The extra degrees of freedom  allow the spherical-UA system to support a  larger number of simultaneous users and reduce the received SNR loss due to interference avoidance. 
}
\end{remark}

\begin{figure}[t]
\begin{center}
\subfigure[Circular UA]{\includegraphics[width=8cm]{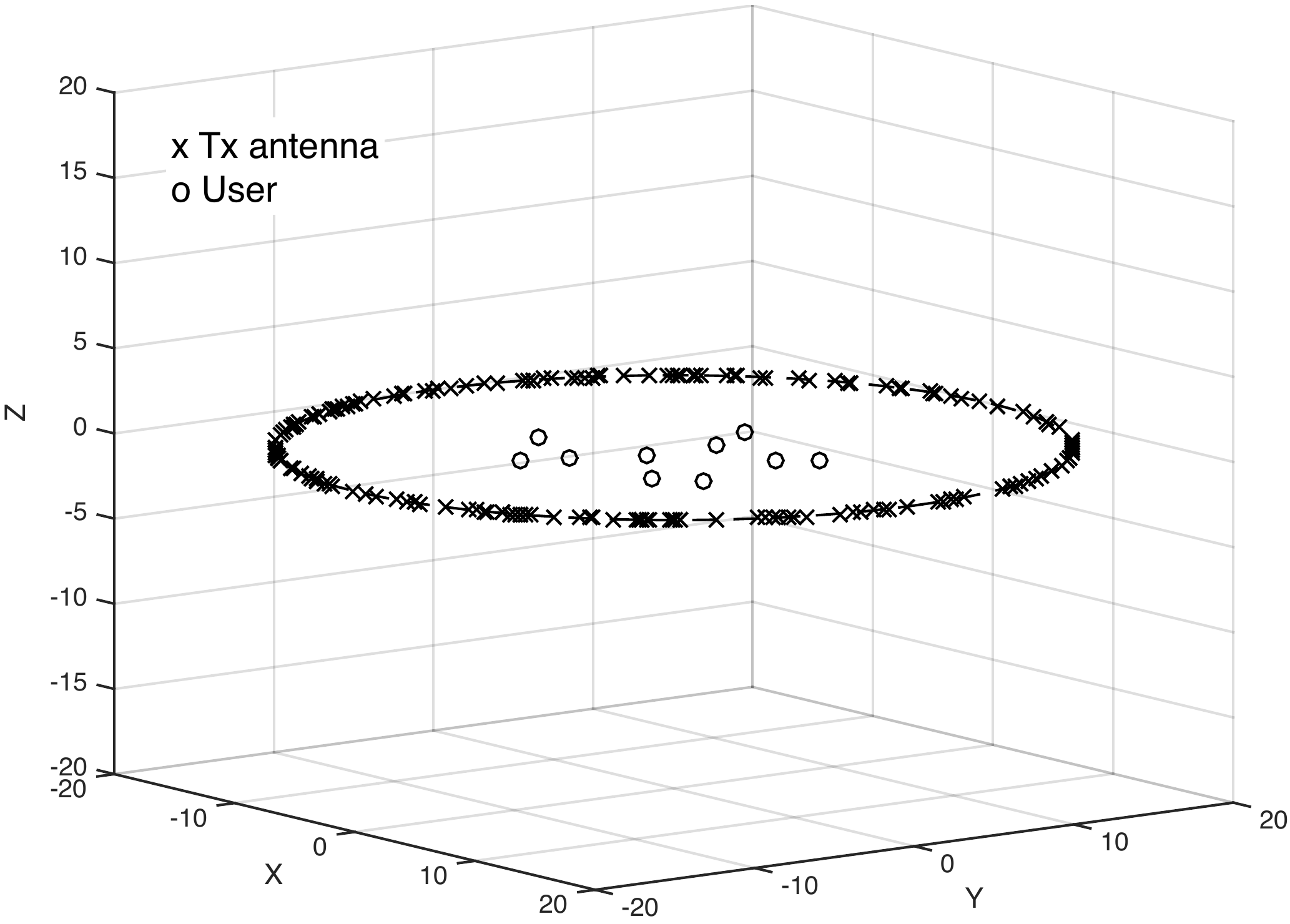}}
\subfigure[Spherical UA]{\includegraphics[width=8cm]{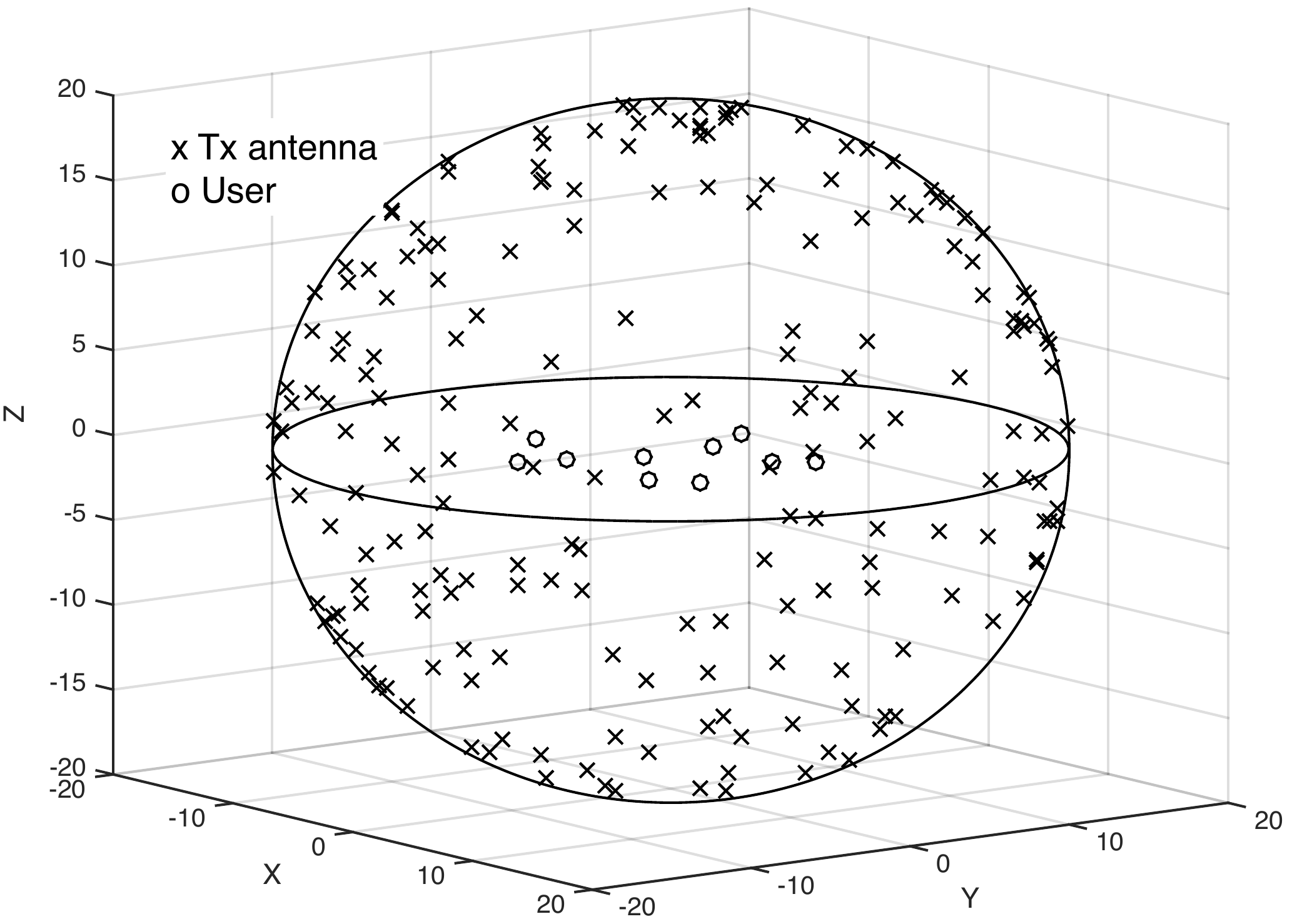}}
\subfigure[Collocated  array]{\includegraphics[width=8cm]{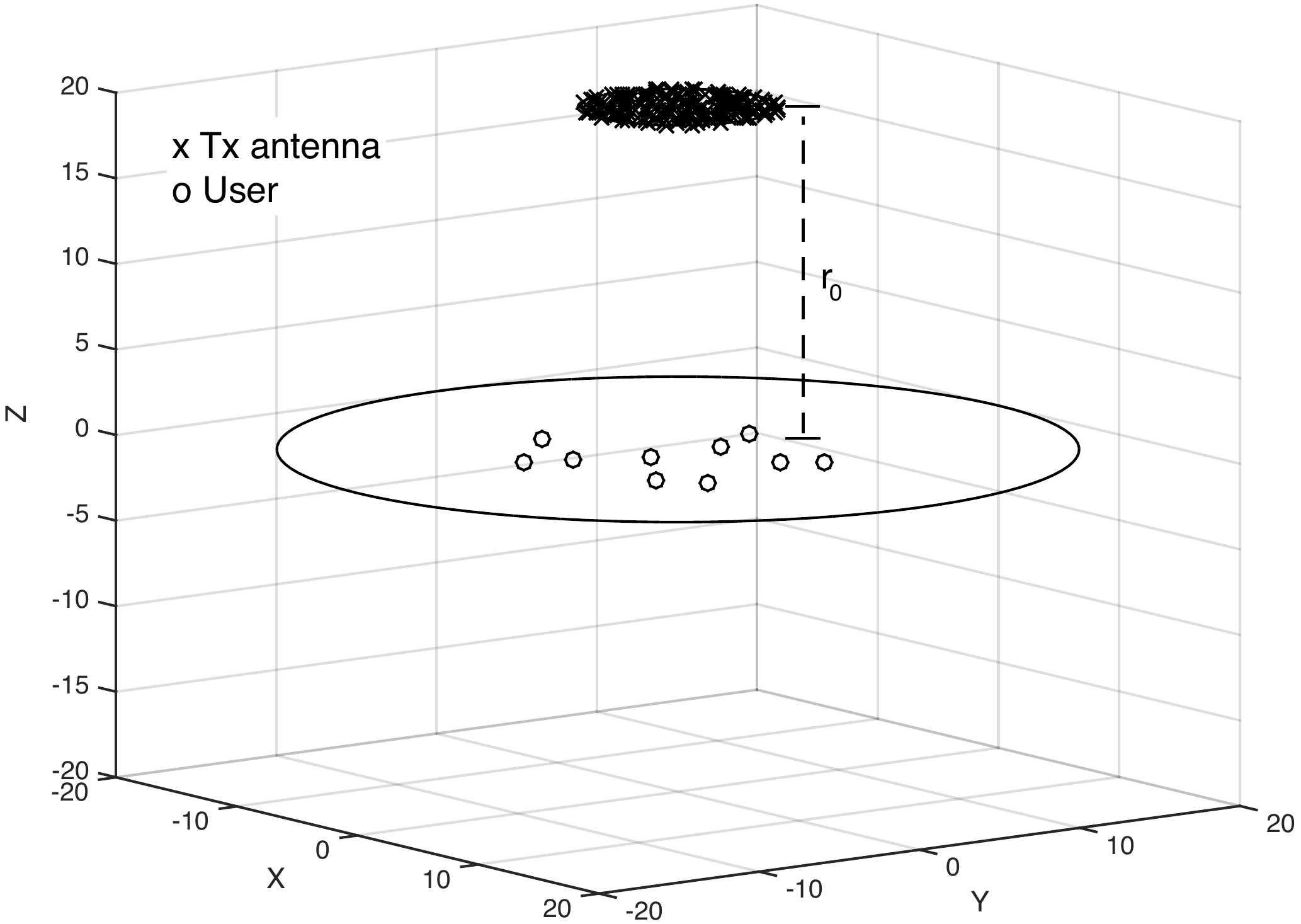}}
\caption{(a) Circular UA with a radius of $r_0 = 20$ m. (b) Spherical UA with the same radius. (c) Collocated array with a radius of $r_d = 5$ m. The number of users is $10$.  }
\label{Fig:UA}
\end{center}
\end{figure}

\section{Simulation   Results}\label{Section:Sim}

In simulation, discrete UAs with finite numbers of antennas  are considered.   As illustrated in Fig.~\ref{Fig:UA}, the UA antennas are uniformly distributed on a circle for the case of  circular UA or a sphere for the case of spherical UA, both of which have the fixed radius $r_0 = 20$ m modeling a small deployment region.  For benchmarking, the conventional \emph{collocated array} is  also included in simulation whose antennas are uniformly distributed in a horizontal disk with a radius denoted as $r_d$ and right above the origin with a vertical distance equal to $r_0$ as shown in Fig.~\ref{Fig:UA}(c). Note that this location and orientation of the collocated UA are found by simulation to yield the best performance among other configurations with the same disk radius  and distance to the origin. The radius of the collocated UA is set as $r_d = 5$ m in Fig.~\ref{Fig:UA} for ease of illustration and $r_d = 1$ m for all other simulation results. In addition,  relaxing Assumption~\ref{AS:CenterUsers},  the users are uniformly distributed in the horizontal disk with a radius of $0.5r_0$ instead of being near the origin. Moreover, the carrier frequency is $2.5$ GHz and the noise variance is $-100$ dBm.

\subsection{Channel Estimation}

\begin{figure}[t]
\begin{center}
\subfigure[Effect of the number of antennas]{\includegraphics[width=8cm]{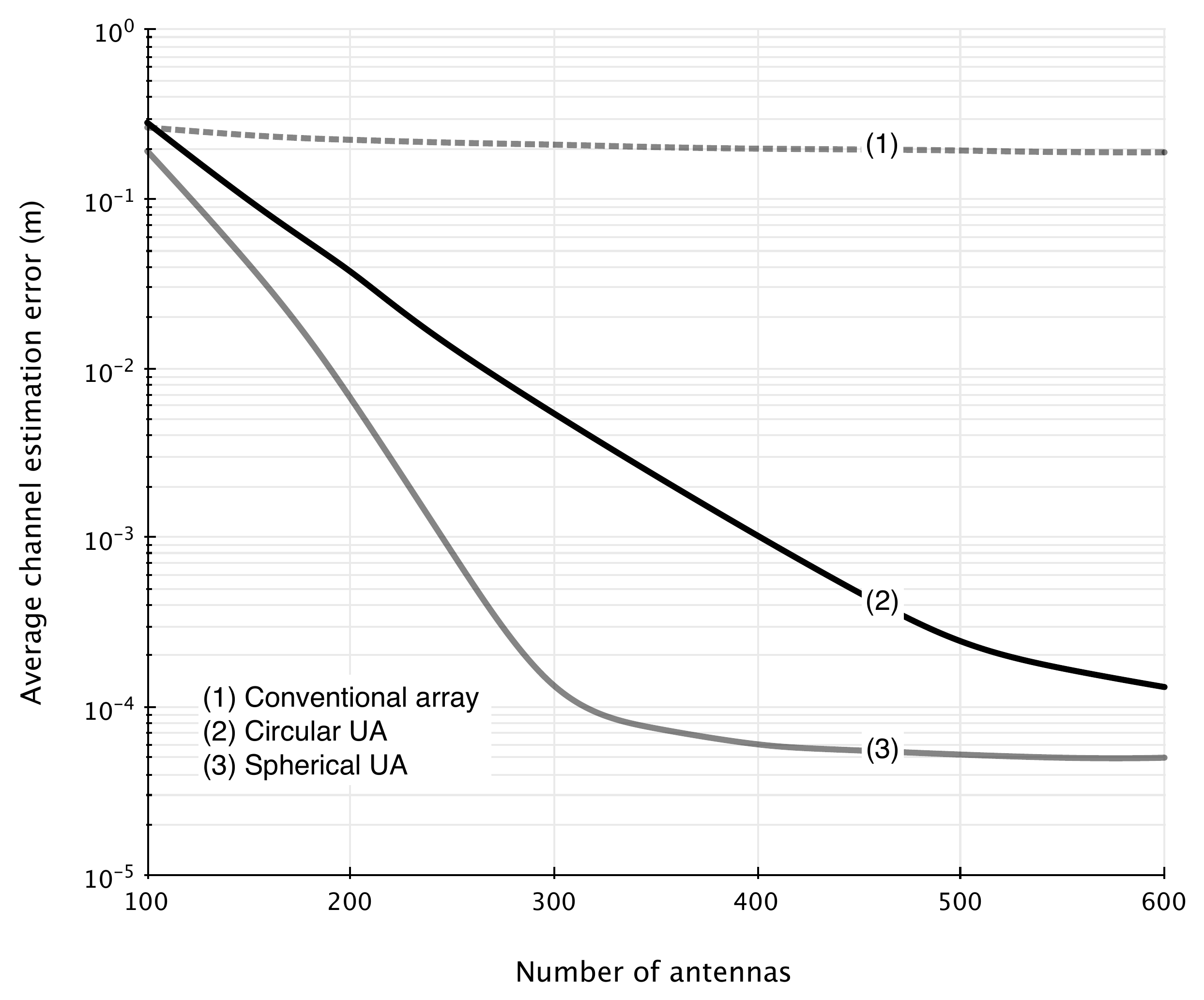}}\hspace{10pt}
\subfigure[Effect of the number of users]{\includegraphics[width=8cm]{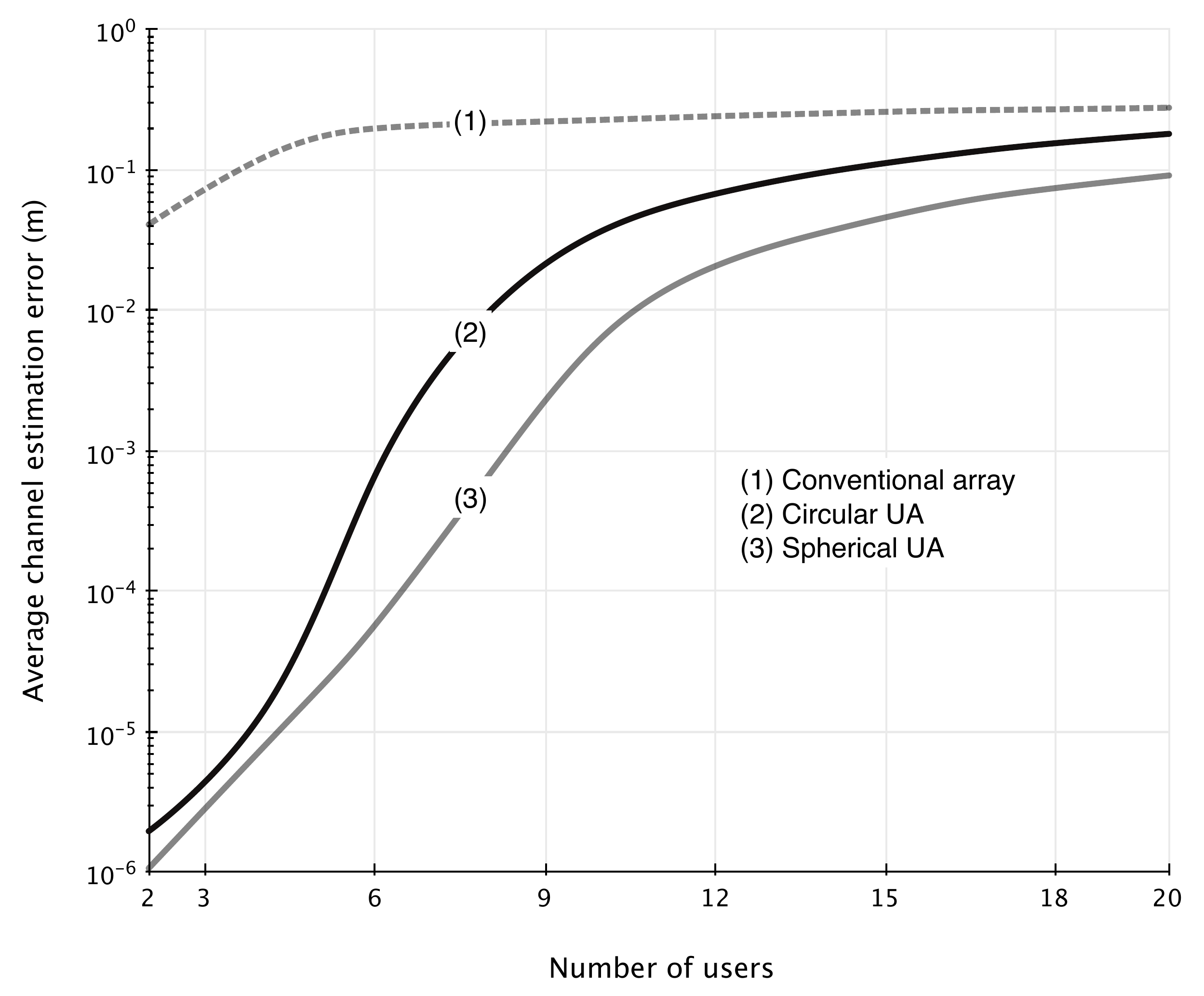}}
\caption{Consider LI-channel estimation with single pilot symbols.  (a) Effect of the number of antennas on channel estimation error for $10$  users.   (b) Effect of the number of users on channel estimation error for $200$ transmit antennas.}
\label{Fig:ChanEst}
\end{center}
\end{figure}

Consider channel estimation using single pilot symbols and algorithms from straightforward extension of those in Sections~\ref{Section:ChanEst} and \ref{Section:ChanEst:Sph} to discrete arrays. The average channel estimation error is defined as the difference between the estimated and actual locations of a typical user as averaged over the random distributions of users and antennas. Fig.~\ref{Fig:ChanEst} displays the curves of average channel estimation error versus the number of antennas and those of average error versus the number of users in two separate sub-figures. Several observations can be made from the curves.  As the  number of antennas increases, the average errors for the circular and spherical UAs both diminishes  rapidly and converges to a small constant corresponding to the continuous arrays analyzed  in the preceding sections. Moreover, the errors grows rapidly as  the  number of users increases.  For relatively small numbers  of users (e.g, fewer than $9$) or large  numbers of antennas (e.g., larger than $200$), the performance of channel estimation for the UAs is much better than that using a collocated UA; the spherical UA outperforms the circular UA. The  conventional array's incapability of accurate LI-channel estimation is mainly due to its confined geometry that is more suitable for estimating signals' angles-of-arrival. For verification, it can observed from Fig.~\ref{Fig:ChanEst} that the average error  for the collocated array is insensitive to the numbers of antennas and users, suggesting that they are not the performance limiting factors.

\subsection{Data Transmission}

\begin{figure}[t]
\begin{center}
\subfigure[Transmit array   with $100$ antennas]{\includegraphics[width=8cm]{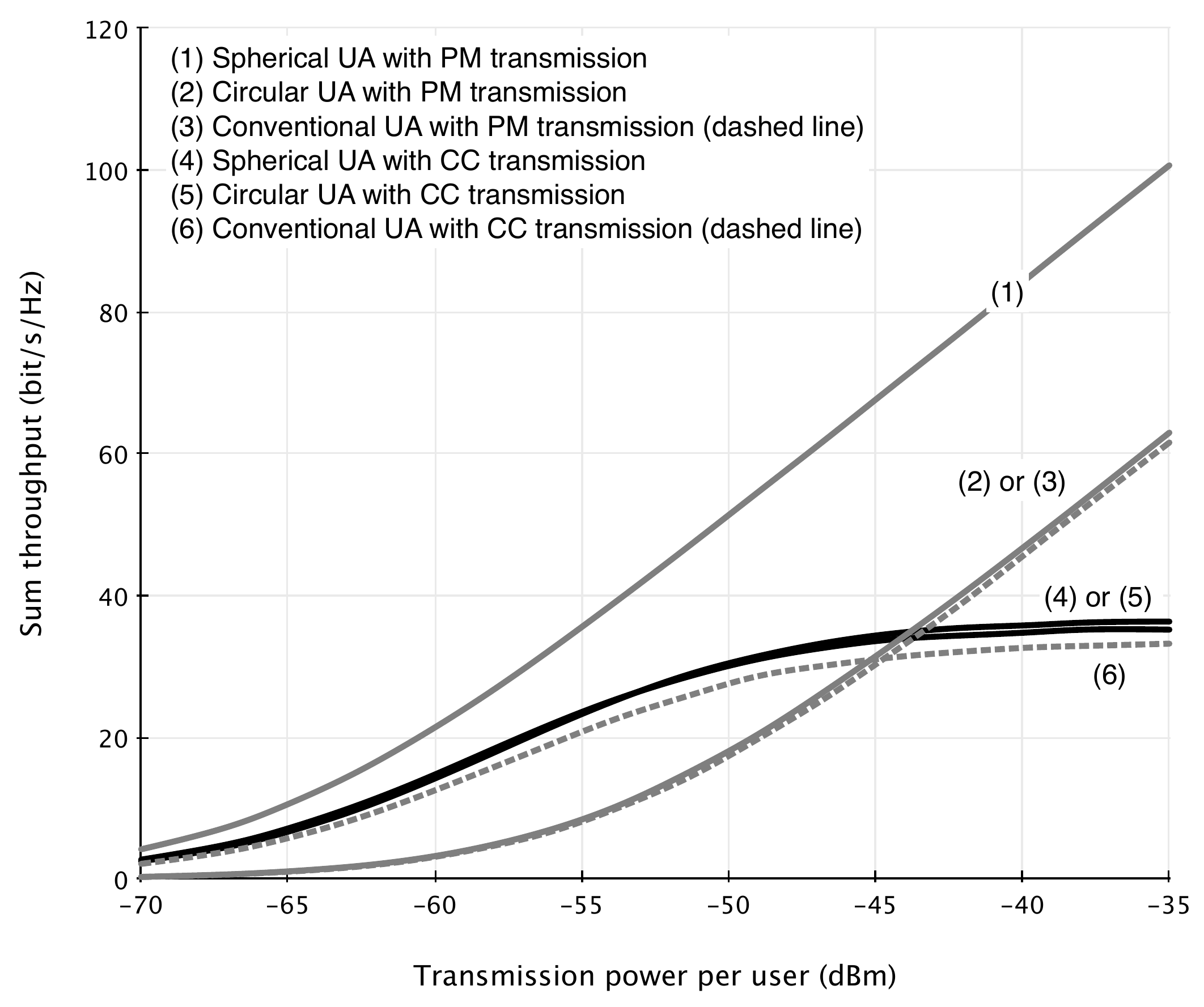}}
\subfigure[Transmit array with  $400$ antennas]{\includegraphics[width=8cm]{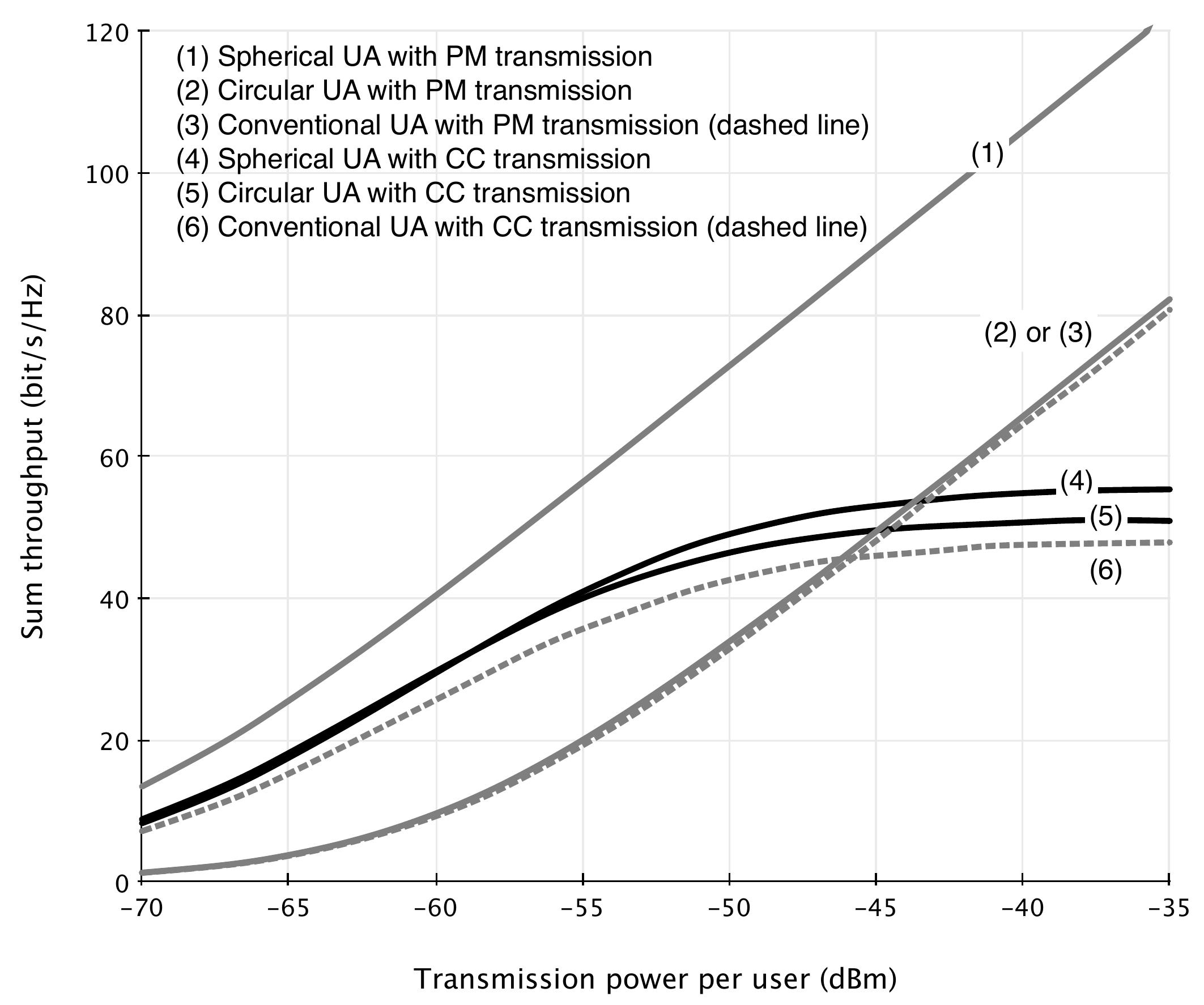}}
\caption{Sum throughput versus transmission power per user for a transmit array with (a) $100$ or (b) $400$ antennas. The number of users is fixed at $10$.   }
\label{Fig:SumCap:TxPwr}
\end{center}
\end{figure}

Consider data transmission assuming perfect channel-state-information at the transmitter.  The curves of sum throughput versus transmission power per user are plotted in Fig.~\ref{Fig:SumCap:TxPwr} with the number of users fixed at $10$. The plots account  for two transmission schemes including  multiuser phase-mode (PM) and single-user channel-conjugate (CC) transmission and different numbers of antennas, namely  $100$ and $400$. Several  observations are made by comparing the curves. First, the combination of spherical UA and PM transmission yields much higher sum throughput than any other combination since  distributing antennas over a larger area allows the spherical UA to generate a much higher number degrees of freedom for avoiding interference and enhancing received signal power compared with other arrays (see Remark~\ref{Re:DoF:Sph}). Second, the sum throughputs for CC transmission using  three types of arrays are comparable and higher than those achieved by PM transmission using the circular UA and collocated UA in the power range of $-70$ to  $-45$ dBm; beyond this range, interference resulting from CC transmission dominates noise, causing the corresponding sum throughputs to saturate. Last, increasing the number of antennas from $100$ to $400$ contributes approximately the same   throughput gain, about  $20$ b/s/Hz, for different combinations of array and transmission scheme.

\begin{figure}[t]
\begin{center}
\subfigure[UA with $100$ antennas]{\includegraphics[width=8cm]{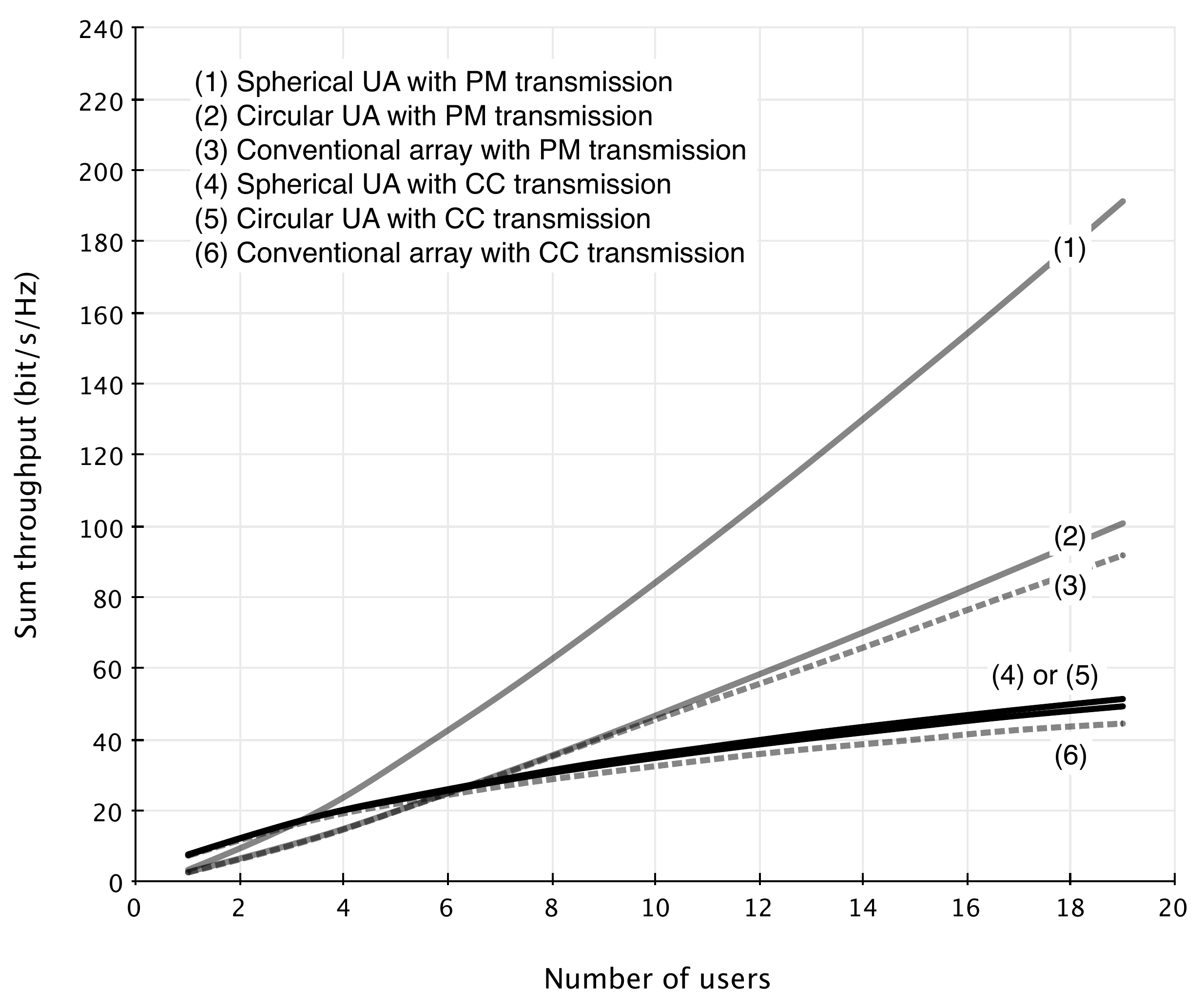}}
\subfigure[UA with $400$ antennas]{\includegraphics[width=8cm]{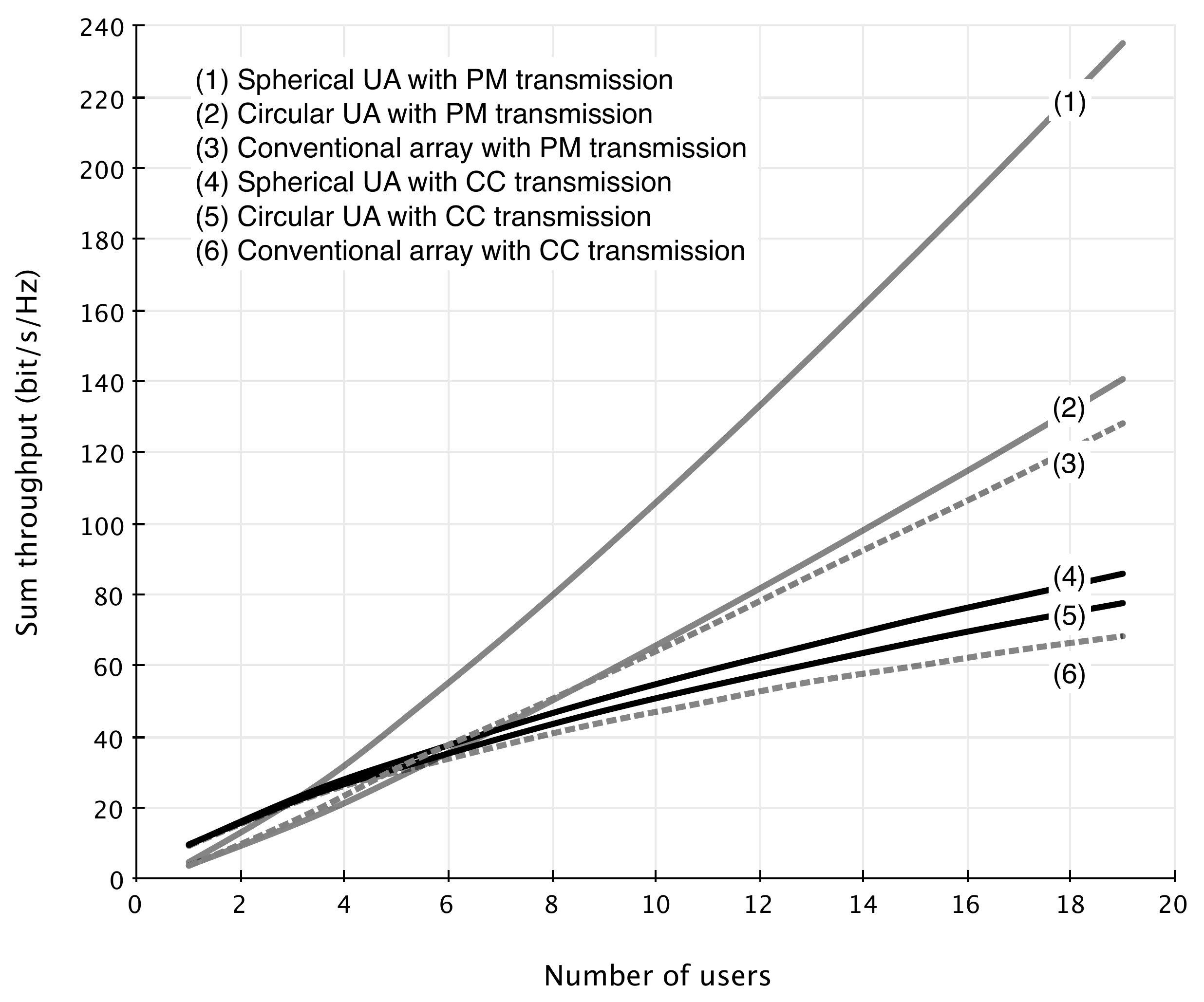}}
\caption{Sum throughput versus number of users for  a transmit array with (a) $100$ or (b) $400$ antennas. The transmission power per  users is fixed at $-40$ dBm.  }
\label{Fig:SumCap:User}
\end{center}
\end{figure}

The curves of sum throughput versus number of users  are plotted in Fig.~\ref{Fig:SumCap:User} with the transmission power per user fixed at  $-40$ dBm. For PM transmission, the sum throughputs are observed to increase approximately \emph{linearly} with the increasing number of users or equivalently the increasing number of simultaneous data streams. Furthermore, the curve corresponding to the spherical UA has a slope  much larger than those for the other types of arrays that are comparable. In contrast, operating in the interference limiting regime,  the sum throughputs for CC transmission using $100$ antennas saturate and are observed to be  insensitive to the increase of the number of users. This issue can be alleviated by deploying more antennas ($400$) that leads to a substantial throughput gain e.g., about $20$ b/s/Hz for the number of users equal to $18$. Nevertheless, the gains for PM transmission are much larger and about $60$ b/s/Hz at  the same number of users. 

\section{Conclusion}\label{Section:Conclusion}
Techniques have been designed for channel estimation and data transmission in the UA system by modeling the UA as a gigantic continuous  circular/spherical array and assuming free-space propagation. It has been shown that the UA enables accurate estimation of multiuser channels even if single pilot symbols are used provided that the user-separation distances  are sufficiently large. If orthogonal pilot sequences are used, channel estimation is found to be always close to perfect.  For single-user data transmission using the UA, inter-user interference can be suppressed by increasing user-separation distances. Alternatively, interference can be nulled at the UA based a novel design of multiuser phase-mode precoders. The resultant number of available degrees of freedom for interference nulling is shown to be proportional to the minimum user-separation distance. Furthermore, the spherical UA provides performance gain compared with the circular counterpart. 

This work opens up several interesting directions for further research. First, the current analysis is based on the model of a continuous circular/spherical UA and targets users near the UA center. Generalizing the model and user locations makes the analysis more challenge and requires the development of new analytical techniques. Second, it is important to address practical issues in the design and analysis of UA communication techniques such as delay and error in message exchange between the UA elements and the presence of sparse scatterers. Furthermore, it is also interesting to design algorithms/protocols for  resource allocation, broadband transmission,  and power control for the UA systems.

\appendices 
\section{Mathematics Preliminary: Bessel Functions and Spherical Harmonics}\label{App:Bessel}
Bessel functions and spherical harmonics   are extensively used  in  the subsequent analysis. In this appendix, the functions are defined and some key properties useful for the analysis  are summarized. 

\subsection{Bessel Functions and Their Properties}
Only  Bessel functions of the first kind are needed in the analysis and referred to  simply as  Bessel functions.    A Bessel function with an integer order $n$ can be defined in an integral form as  \cite{arfken_2005_mathematical_physics}:
\begin{equation}
J_n(x) = \frac{1}{2\pi}\int_0^{2\pi} e^{j(x\sin \phi - n\phi)} d\phi.   \label{Eq:Bessel:Def}
\end{equation}

Several useful properties of  Bessel functions are described as follows. 

\begin{enumerate}
\item [(B1)] The Jacobi-Anger expansion decomposes an  exponential function of a trigonometric function into its harmonics as follows \cite{arfken_2005_mathematical_physics}:
\begin{equation}
e^{jx\cos\phi} = \sum_{n=-\infty}^\infty j^n J_n(x) e^{j k \phi}. 
\end{equation}

\item [(B2)] Addition Theorem I   \cite[(6.61)]{andrews_1992_special}: 
\begin{equation}
  J_n(R) e^{jn\omega} = \sum_{k=-\infty}^\infty J_k(a)J_{n+k}(b)e^{jk\phi}\nn
\end{equation}
where $R=\sqrt{a^2+b^2-2ab\cos\phi}$ and $\sin\omega  = (a/R)\sin\phi$.

\item [(B3)] Addition Theorem II rewritten from     \cite[(6.62)]{andrews_1992_special}: 
\begin{equation}
\mathrm{sinc}(R) = \frac{\pi}{2}  \sum_{n=0}^\infty \l(2n + 1\r)\frac{J_{n+\frac{1}{2}}(a)J_{n+\frac{1}{2}}(b)}{\sqrt{ab}}P_n(\cos \phi)\nn
\end{equation}
where $P_n(x)$ with $x\in [-1, 1]$ is the Legendre polynomial defined as 
\begin{equation}\label{Eq:Legendre}
P_n(x) = \frac{1}{2^n n !}\frac{d^n}{dx^n}(x^2 - 1)^n. 
\end{equation}

\item [(B4)]  The Bessel function $J_0(x)$  can be upper bounded as \cite{landau_2000_bound_besselfun} 
\begin{equation}
J_0(x) \leqslant  \nu x^{-\frac{1}{3}}
\end{equation}
where the constant $\nu = 0.7857\cdots$. 

\item [(B5)] Given $0 < z \leq 1$,   a Bessel function with a high order satisfies  \cite[$9.3.5$ and $9.3.6$]{AbramowitzBook}\footnote{Two functions $f$ and $g$ are \emph{asymptotic equivalent}, denoted as $f(x) \sim g(x)$, if $\lim_{x\rightarrow\infty} f(x)/g(x) = 1$.} 
\begin{equation}
J_n(z n ) \sim \frac{c(z)}{n^{\frac{1}{3}}}, \qquad n \rightarrow \infty
\end{equation}
where $c(z)$ is a positive constant whose value depends only  on $z$.  Consequently, for $x\gg 1$ and $|n| \geqslant x$, $J_n(x) \approx 0$  \cite{PoonTse:DoFMultiAntennaChannels:2005}.

\item[(B6)] Gegenbauer's generalization of  Poisson's integral   \cite[$10.1.14$]{AbramowitzBook}: 
\begin{equation}\label{Eq:Poisson:Int}
   \sqrt{\frac{\pi}{2x}} J_{n+\frac{1}{2}}(x) = \frac{1}{2}(-j)^n \int_{-1}^1 e^{jx\tau}P_n(\tau)d\tau
\end{equation}
where $P_n(x)$ is the Legendre polynomial defined in \eqref{Eq:Legendre}. 
\end{enumerate}

\subsection{Spherical Harmonics  and Their Properties}

The  spherical harmonic functions denoted as   $\{Y^m_{\ell}(\phi, \theta)\}$ with integer indices $\ell = 0, 1, \cdots$ and $-\ell \leqslant m \leqslant \ell$ are  defined as \cite{arfken_2005_mathematical_physics}: 
\begin{equation}\label{Eq:Sph:Basis}
  Y^m_{\ell}(\phi, \theta)  =  \sqrt{\frac{2m+1}{4\pi}\frac{(\ell-m)!}{(\ell+m)!}}P^m_{\ell}(\cos \theta)e^{jm\phi}, \qquad \theta\in [0, \pi], \phi \in [0, 2\pi]
\end{equation}
where $P^m_{\ell}(\cos \theta)$ represents   the  \emph{associated Legendre polynomial} defined as 
\begin{equation}\label{Eq:Legendre:Assoc}
P^m_{\ell}(x) = \frac{(-1)^m}{2^\ell \ell !}\l(1 - x^2\r)^{\frac{m}{2}} \frac{d^{m+\ell}}{dx^{m+\ell}} \l(x^2 - 1\r)^\ell, \qquad -1 \leq x \leq 1. 
\end{equation}
Note that  $P^m_{\ell}(\cos \phi)$ with $m  = 0$ reduces to the Legendre polynomial in \eqref{Eq:Legendre}.  

Several useful properties of the spherical harmonics  are described as follows. 
\begin{enumerate}
\item [(S1)] The functions $\{Y^m_{\ell}(\phi, \theta)\}$ are orthonormal over the spherical surface: 
\begin{equation}\label{Eq:Orthogonal}
\int_{\phi = 0}^{2\pi}\int_{\theta = 0}^\pi   [(Y_{\ell}^m)(\phi, \theta)]^*  Y_{\ell'}^{m'}(\phi, \theta) \sin\theta \ d\theta d\phi = \delta_{m, m'}\delta_{\ell, \ell'} 
\end{equation}
where $\delta_{m, m'} $ is equal to $1$ if $m = m'$ and $0$ otherwise.

\item[(S2)] Funk-Hecke Theorem  \cite[Theorem~$3$]{Gumerov:FastMultipoleHelmholtzEq:2005}: Let $\psi_u(\phi, \theta)$ denote the angle between the vectors $X_u = (r_u, \phi_u, \theta_u)$ and $(1, \phi, \theta)\in \mathds{R}^3$. The Laplace series of a function $w(\cos \psi_u(\phi, \theta))$ is given as
\begin{equation}\nn
w(\cos \psi_u(\phi, \theta)) = \sum_{\ell=0}^\infty \sum_{m=-\ell}^\ell c_{\ell}^m(\phi_u, \theta_u) Y^m_{\ell}(\phi, \theta)
\end{equation}
where the coefficient $c_{\ell}^m(\phi_u, \theta_u) = c_\ell Y^m_{\ell}(\phi_u, \theta_u)$ 
with 
\begin{equation}
  c_\ell = 2\pi \int_{-1}^1  w(\tau ) P_\ell(\tau)d\tau.  \nn
\end{equation}

\item[(S3)] Spherical Harmonic Addition Theorem \cite[($16.57$)]{arfken_2005_mathematical_physics}: 
\begin{equation}\label{Eq:Sph:Add}
\sum_{m=-\ell}^\ell [Y^m_{\ell}(A)]^* Y^m_{\ell}(X_u) = \frac{2\ell+1}{4\pi}P_\ell\left(\cos\psi_u(A)\right) 
\end{equation}
where $A, X_u \in \mathds{R}^3$ and $\psi_u(A)$ denotes their separation angle. 

\end{enumerate}

\section{Proofs of Lemmas}\label{App:Proofs}
\subsection{Proof of Lemma~\ref{Lem:Fourier}}\label{Lem:Fourier:Proof}

By substituting \eqref{Eq:Training:Sig} and $\phi_u(A) = \phi_u - \phi$ into \eqref{Eq:Fourier}, 
\begin{align}
Q_k  &=   \frac{\lambda e^{-j \frac{2\pi}{\lambda}r_0}}{4\pi r_0} \sum_{u=1}^U \int_0^{2\pi}    e^{j\frac{2\pi }{\lambda}r_u\cos(\varphi_u-\phi)}e^{jk\phi} d\phi + \frac{1}{2\pi}\int_0^{2\pi} z(\phi) e^{jk\phi}d\phi + \frac{o}{r_0}. \label{Eq:Spectrum}
\end{align}
The noise term can be written as 
\begin{equation}\label{Eq:Noise}
\frac{1}{2\pi}\int_0^{2\pi} z(\phi) e^{jk\phi}d\phi = \lim_{N\rightarrow\infty}\frac{1}{N}\sum_{n=1}^N z\l(\frac{2\pi n}{N}\r) e^{jk\frac{2\pi n}{N}}. 
\end{equation}
For ease of notation, define $\tilde{z}_n= z\l(\frac{2\pi n}{N}\r) e^{jk\frac{2\pi n}{N}}$. Since $z_0, z_1, \cdots, \tilde{z}_N$ is an i.i.d. sequence of $\mathcal{CN}(0, \sigma^2)$ random variables under Assumption~\ref{AS:White}, by applying the law of large numbers, it follows from  \eqref{Eq:Noise} that
\begin{equation}\label{Eq:Noise:a}
\frac{1}{2\pi}\int_0^{2\pi} z(\phi) e^{jk\phi}d\phi = 0, \qquad \text{a.s.}
\end{equation}
Next, based on the Jacobi-Anger expansion in Property (B1) of Bessel functions in Appendix~\ref{App:Bessel}, the first exponential term in \eqref{Eq:Noise} can be decomposed as 
\begin{equation}\label{Eq:Jacobi-Anger}
e^{j\frac{2\pi }{\lambda}r_u\cos(\varphi_u-\phi)} = \sum_{n=0}^\infty j^{n}J_{n}\left(\tfrac{2\pi }{\lambda} r_u \right) e^{j n(\varphi_u-\phi)}. 
\end{equation}
By substituting  \eqref{Eq:Noise:a} and \eqref{Eq:Jacobi-Anger} into \eqref{Eq:Spectrum}, it can be obtained that 
\begin{equation}
Q_k  =   \frac{\lambda e^{-j \frac{2\pi}{\lambda}r_0}}{4\pi r_0}  \sum_{u=1}^U    \sum_{n=0}^\infty j^{n}J_{n}\left(\tfrac{2\pi }{\lambda}r_u\right) \int_0^{2\pi} e^{jn(\varphi_u-\theta)}e^{jk\theta} d\theta  + \frac{o}{r_0}, \qquad \text{a.s.} \label{Eq:Spectrum:a}
\end{equation}
Based on the following equality 
\begin{equation}\label{equ:exponentials_integration}
  \frac{1}{2\pi}\int_0^{2\pi}  e^{j(k-n)\theta} d\theta = \delta_{k, n}, \nn
\end{equation}
the desired result follows from \eqref{Eq:Spectrum:a}. \hfill\qed

\subsection{Proof of Lemma~\ref{Lem:CC:Field}}\label{App:CC:Field}
Using trigonometric identities, it can be obtained that
\begin{align}
r_u \cos(\varphi_u-\theta)-r_k\cos(\varphi_k-\theta) & = |X_u - X_k| \l(\cos \beta_{u, k} \cos\theta - \sin\beta_{u, k}\sin\theta\r)\nn\\
&= |X_u - X_k| \cos (\beta_{u, k}+\theta)  \nn
\end{align}
where the angle $\beta_{u, k}$ is defined by 
\begin{equation}
\tan \beta_{u, k} = \frac{r_u\cos(\phi_u) - r_k\cos(\phi_k)}{r_u\sin(\phi_u) - r_k\sin(\phi_k)}. 
\end{equation}
It follows that
\begin{align}
\frac{1}{2\pi}  \int_0^{2\pi}e^{j\frac{2\pi}{\lambda}\left(r_u \cos(\varphi_u-\theta)-r_k\cos(\varphi_k-\theta)\right)} \ d\theta  &= \frac{1}{2\pi}  \int_0^{2\pi}e^{j\frac{2\pi}{\lambda}|X_u - X_k| \cos (\beta_{u, k}+\theta) } \ d\theta\nn\\
&= \frac{1}{2\pi}  \int_0^{2\pi}e^{j\frac{2\pi}{\lambda}|X_u - X_k| \sin\l(\beta_{u, k}+\theta + \tfrac{\pi}{2}\r)} \ d\theta. \nn
\end{align}
Using the definition of  Bessel function in \eqref{Eq:Bessel:Def}, 
\begin{equation}
\frac{1}{2\pi}  \int_0^{2\pi}e^{j\frac{2\pi}{\lambda}\left(r_u \cos(\varphi_u-\theta)-r_k\cos(\varphi_k-\theta)\right) } \ d\theta =  j e^{j \beta_{u, k}} J_m\l(\tfrac{2\pi}{\lambda}|X_u - X_k|\r). 
 \label{Eq:Trigo}
\end{equation}
Substituting \eqref{Eq:Trigo} into \eqref{Eq:Field:Dist:cir:a} gives 
\emph{\begin{equation}
g(X_k\mid X_u) = \sqrt{\frac{P_{\textrm{t}}}{2 r_0}} j e^{j \beta_{u, k}} J_0\l(\tfrac{2\pi}{\lambda}|X_u - X_k|\r)  + \frac{o}{\sqrt{r_0}}. \nn
\end{equation}}
The  desired result follows.   \hfill\qed

\subsection{Proof of Lemma~\ref{Lem:Sph:Expan}}\label{Lem:Sph:Expan:Proof}
By substitution of the training signal in \eqref{Eq:Training:Sig}, its Laplace coefficients defined in \eqref{Eq:Q:Sph} are obtained as 
\begin{align}
Q_{\ell}^m &= \frac{\eta}{r_0}\sum_{u=1}^U\int_{\theta=0}^{2\pi} \int_{\psi=0}^\pi e^{j\frac{2\pi}{\lambda}r_u\cos \psi_u }Y^m_{\ell}(\phi, \theta) \sin\phi d\phi d\theta+\nn\\
&\qquad \qquad  \int_{\theta=0}^{2\pi} \int_{\psi=0}^\pi z(\phi, \theta) \sin\phi d\phi d\theta + \frac{o}{r_0}\nn\\
&= \frac{\eta}{r_0}\sum_{u=1}^U\int_{\theta=0}^{2\pi} \int_{\psi=0}^\pi e^{j\frac{2\pi}{\lambda}r_u\cos \psi_u }Y^m_{\ell}(\phi, \theta) \sin\phi d\phi d\theta + \frac{o}{r_0}, \qquad   \text{a.s.}\label{Eq:Q:Sph:App}
\end{align}
where the vanishment of noise follows similar analysis as in the proof for Lemma~\ref{Lem:Fourier}.  Using the Funk-Hecke Theorem in (S$2$) , the exponential term in the last equation can be also expanded into a Laplace series as 
\begin{equation}\label{Eq:Sph:Exp}
  e^{j\frac{2\pi}{\lambda} r_u\cos \psi_u } = \sum_{\ell=0}^\infty \sum_{m=-\ell}^\ell c_\ell Y^m_{\ell}(\phi_u, \theta_u) Y^m_{\ell}(\psi, \theta)
\end{equation}
where   
\begin{equation}
  c_\ell = 2\pi \int_{-1}^1  e^{j\frac{2\pi}{\lambda} r_u \tau } P_\ell(\tau)d\tau.   \label{Eq:Cn}
\end{equation}
Using Property (B$6$) in Appendix~\ref{App:Bessel}, it follows from \eqref{Eq:Cn}  that
\begin{equation}\label{Eq:Cn:more}
  c_\ell(r_u) =   \frac{(2\pi)^{\frac{3}{2}} j^\ell J_{\ell+\frac{1}{2}}\left(\frac{2\pi}{\lambda}r_u\right)}{\left(\frac{2\pi}{\lambda}r_u\right)^{\frac{1}{2}}}  Y^m_{\ell}(\phi_u, \theta_u). 
\end{equation}
Combining \eqref{Eq:Sph:Exp} and \eqref{Eq:Cn:more} gives the desired result. \hfill\qed

\subsection{Proof of Lemma~\ref{Lem:J:Spy}}\label{App:J:Spy:Proof}
Since $\psi_u$ is the angle between two unit vectors with spherical coordinates $(1, \phi, \theta)$ and $(1, \phi_u, \theta_u)$, 
\begin{equation}
  \cos \psi_u =  \sin\phi\cos \theta\sin\phi_u\cos\theta_u + \sin\phi\sin \theta\sin\phi_u\sin\theta_u + \cos\phi\cos \phi_u.
\end{equation}
It follows that 
\begin{align}
  r_u \cos \psi_u - r_k\cos \psi_k & =  \sin\phi\cos \theta(r_u\sin\phi_u\cos\theta_u - r_k\sin\phi_k\cos\theta_k) + \nn\\
  &\qquad \sin\phi\sin \theta(r_u \sin\phi_u\sin\theta_u - r_k \sin\phi_k\sin\theta_k)+\nn\\
  &\qquad   \cos\phi(r_u\cos \phi_u - r_k\cos \phi_k).\label{Eq:PhaseDiff:Sph}
\end{align}
To rewrite   the right-hand side of \eqref{Eq:PhaseDiff:Sph} in a compact form, define the spherical coordinates $(r_{u, k}, \phi_{u, k}, \theta_{u, k})\in\mathds{R}^3$ such that 
\begin{align}
r_{u, k}^2 &= (r_u\sin\phi_u\cos\theta_u - r_k\sin\phi_k\cos\theta_k)^2 + (r_u \sin\phi_u\sin\theta_u - r_k \sin\phi_k\sin\theta_k)^2 + \nn\\
&\qquad (r_u\cos \phi_u - r_k\cos \phi_k)^2\label{Eq:r_uk:Def}
\end{align}
and the angles $\phi_{u, k}$ and  $\theta_{u, k}$ as in the lemma statement. Moreover, let  $\psi_{u, k}$ denote the angle between $(r_0, \phi, \theta)$ and $(r_{u, k}, \phi_{u, k}, \theta_{u, k})$. Using these definitions, \eqref{Eq:PhaseDiff:Sph} can be reduced to
\begin{align}
  r_u \cos \psi_u - r_k\cos \psi_k & =    r_{u, k} \cos \psi_{u, k} 
\end{align}
It can be obtained from \eqref{Eq:r_uk:Def} that $r_{u, k}^2 = |X_u - X_k|^2$. Thus, 
\begin{equation}
  r_u \cos \psi_u - r_k\cos \psi_k =    |X_u - X_k| \cos \psi_{u, k}. 
\end{equation}
Substituting this result into \eqref{Eq:J:Sph:Def} yields 
\begin{equation}
\mathcal{J}_{u, k, m, \ell} = \iint e^{j\tfrac{2\pi}{\lambda} |X_u - X_k| \cos \psi_{u, k}} Y^m_{\ell}(\phi, \theta) \sin\phi \ d\phi \ d\theta.    \label{Eq:J:Sph:Proof} 
\end{equation}
Applying Lemma~\ref{Lem:Sph:Expan} gives the desired result. \hfill\qed

\bibliographystyle{ieeetr}

\end{document}